\documentclass[amsmath,amssymb,pra,superscriptaddress]{revtex4}
\usepackage{graphicx}
\usepackage{dcolumn}
\usepackage{bm}
\usepackage{subfigure}
\usepackage{booktabs}
\usepackage{color}
\newcounter{one}
\newcommand{\bra}[1]{\langle #1 |}
\newcommand{\ket}[1]{| #1 \rangle}

\newcommand{\braket}[2]{\langle #1 | #2 \rangle}

\newcommand{\affA}{
Department of Physics, The University of Tokyo, 4-6-1 Komaba, Meguro, Tokyo 153-8505, Japan
}

\begin{document}

\title{\textbf{Microscopic analysis of\\ the microscopic reversibility in quantum systems}}

\author{Tatsuro Kawamoto}
\affiliation{\affA}
\date{\today}

\begin{abstract}
We investigate the robustness of the microscopic reversibility in open quantum systems which is discussed by 
Monnai [arXiv:1106.1982 (2011)].
%Ref.~\cite{2011arXiv1106.1982M}.  
We derive an exact relation between the forward transition probability and the reversed transition probability in the case of a general measurement basis.  
We show that the microscopic reversibility acquires some corrections in general and discuss the physical meaning of the corrections.
Under certain processes, some of the correction terms vanish and we numerically confirmed that 
the remaining correction term becomes negligible; 
for such processes, the microscopic reversibility almost holds even when the local system cannot be regarded as macroscopic.
\end{abstract}

\maketitle

\section{Introduction}\label{Intro}
Understanding of non-equilibrium quantum dynamics has been eagerly pursued since many experiments are done under non-equilibrium situations and should be also treated quantum mechanically. 
While there are many physical quantities and relations to characterize the properties of the equilibrium systems, 
very few are known for the systems out of equilibrium.  
The linear response theory, which describes the non-equilibrium quantities in terms of the equilibrium quantities, is a powerful tool to investigate such systems.
It is, however, restricted to the case where the systems are close to equilibrium.  
In order to describe the behaviors of strongly non-equilibrium systems, it may be required to find some relations which contain detailed information of the dynamics.

For exploring the character of the strongly non-equilibrium dynamics of quantum systems,
symmetry relations for non-equilibrium states such as fluctuation theorems are of great importance \cite{PhysRevLett.100.230404,2010PhRvL.105n0601C,PhysRevLett.102.210401,1742-5468-2008-10-P10023,2000cond.mat..7360K,PhysRevE.72.027102,2003cond.mat..8337M,Talkner:2009fr,2000cond.mat..9244T}.
In particular, significant attentions are paid to the relation between the transition probabilities of forward processes and the transition probabilities of the corresponding reversed processes \cite{2010PhRvL.105n0601C,PhysRevLett.102.210401,2000cond.mat..7360K,PhysRevE.72.027102,2003cond.mat..8337M,Talkner:2009fr,2000cond.mat..9244T}.
Such relations are appealing particularly because they are not restricted to the close-to-equilibrium states.
As the forward process of the fluctuation theorems, it is typical to consider a driven system which is in the thermal equilibrium state at $t=0$ and then is controlled by a time-dependent external parameter $\lambda(t)$. 
We consider the case where the states at the beginning and the ending of the process are measured, and thus the states are projected.
We can interpret that the state hopped from the one to the other; 
the transition probability of the forward process is defined as the probability that such a transition occurs.  
%One way is to define it as the probability that a state evolves along a certain trajectory in the state space. 
%
%We explore this kind of relation from the viewpoint of quantum operation on a local system in a reservoir.
%
For the reversed process, it is common to consider the following 
in quantum systems \cite{2010PhRvL.105n0601C,PhysRevLett.102.210401,2000cond.mat..7360K,Talkner:2009fr,2000cond.mat..9244T}.  
% and classical systems \cite{PhysRevE.60.2721,Harris:2007fj,PhysRevLett.95.040602,2008ARPC...59..603S}. 
%
The initial density matrix of the reversed process is set to be the same as that of the forward process but with the value of the external parameter $\lambda(T)$,  where $T$ is the final moment of the forward process.  
Then, the system is driven with a time-reversed protocol $\lambda(T-t)$; 
the transition probability of the reversed process is defined as the probability that the opposite transition occurs compared to the transition of the forward process. 

They are the typical settings for the study of fluctuation theorems.  
From the viewpoint of quantum operation, however, it may be natural to set the time-evolved (generally non-equilibrium) state as the initial density matrix of the reversed process.  

As another point for the study of open quantum systems, the choice of the measurement process is especially significant; 
measurement processes on a reservoir \cite{1742-5468-2008-10-P10023} or both on a local system and a reservoir 
\cite{2010PhRvL.105n0601C,PhysRevLett.102.210401,2000cond.mat..7360K,Talkner:2009fr} are often considered.
Nevertheless, sometimes it is more natural to consider the measurement solely on the local system; 
in the present paper, we consider such a case.
There are fluctuation theorems of open quantum systems which are written solely in the terms of local system though they do not take into account of the measurement process in order to discuss the forward and the reversed processes 
\cite{EspositoMukamel,KawamotoHatano}.

The above two points were taken into account in a recent study by Monnai \cite{2011arXiv1106.1982M}.  
The study pointed out the significance of the microscopic reversibility in open quantum systems 
as a kind of symmetry relation similar to but different from the fluctuation theorems \cite{2011arXiv1106.1982M}.
It defines the reversed process as the one from the time-evolved state and considers the measurements on the local system only.

The discussion in Ref.~\cite{2011arXiv1106.1982M} for open quantum systems is as follows.
The total Hamiltonian consists of the Hamiltonian of a local system $\hat{H}_{\mathrm{s}}(\lambda(t))$ which is controled by external forces with the parameter $\lambda(t)$, 
the Hamiltonian of a reservoir $\hat{H}_{\mathrm{r}}$, and the Hamiltonian of coupling between them $\hat{H}_{\mathrm{c}}$, i.e.\,,
\begin{align}
\hat{H}_{\mathrm{tot}}(t)  = \hat{H}_{\mathrm{s}}(\lambda(t)) + \hat{H}_{\mathrm{r}} + \hat{H}_{\mathrm{c}}.  
\end{align}
Let us consider the process where we measure the states of the local system at $t=0$ and $t=T$.  
Thoughout this paper, we employ the Schr\"{o}dinger picture and only consider the projection measurement as the measurement protocol.
The measurement basis at $t=0$ can be different from the one at $t=T$.  
We refer to the measured states as $\ket{n(0)}$ and $\ket{m(T)}$ and to the probability of such a transition as $p_{\mathrm{F}}(\ket{n(0)}\rightarrow\ket{m(T)})$.
Note that the time variables of $\ket{n(0)}$ and $\ket{m(T)}$ merely indicate the moments that the measurements are done along the forward process; 
they do not mean that those measurement bases are time dependent. 
Next, the reversed process is defined as follows; 
as the initial state, 
we prepare the state $\hat{\rho}(T)$ that evolved from $t=0$ to $t=T$ \textit{without} the measurement at $t=0$, 
and then drive the system from $t=T$ to $t=2T$ with the time-reversed protocol $\lambda(2T-t)$. 
The reversed transition probability is defined to be the probability of observing $\hat{\Theta}\ket{m(T)}$ at $t=T$ and $\hat{\Theta}\ket{n(0)}$ at $t=2T$ under the time-reversed process, where $\hat{\Theta}$ is the time-reversal operator.
We refer to such a transition probability as $p_{\mathrm{R}}(\hat{\Theta}\ket{m(T)}\rightarrow\hat{\Theta}\ket{n(0)})$.
Monnai then showed the equality \cite{2011arXiv1106.1982M} 
\begin{align}
p_{\mathrm{F}}(\ket{n(0)}\rightarrow\ket{m(T)})
=p_{\mathrm{R}}(\hat{\Theta}\ket{m(T)}\rightarrow\hat{\Theta}\ket{n(0)}) \label{microrev}
\end{align}
under the following conditions:
(i) the total system is a product state at $t=0$; 
(ii) %the coupling between the local system and the reservoir is weak 
the local system is macroscopic, so that the contribution from the coupling Hamiltonian $\hat{H}_{\mathrm{c}}$ is extremely small compared to the ones from the local system $\hat{H}_{\mathrm{s}}$ and the reservoir $\hat{H}_{\mathrm{r}}$, and we thereby have 
$\hat{U}(T) ( \sqrt{\hat{\rho}_{\mathrm{s}}(0)} \otimes \sqrt{\hat{\rho}_{\mathrm{r}}(0)} ) \hat{U}^{\dagger}(T) 
\simeq \sqrt{\hat{\rho}_{\mathrm{s}}(T)} \otimes \sqrt{ \hat{\rho}_{\mathrm{r}}(0) }$, 
where $\hat{\rho}_{\mathrm{s}}$ and $\hat{\rho}_{\mathrm{r}}$ are the density matrices of the local system and the reservoir, respectively; 
and 
(iii) the measured states at $t=0$ and $t=T$ are the eigenstates of the density matrix of the local system.
We call the equality (\ref{microrev}) the microscopic reversibility in open quantum systems.
Note that it is a relation about the local system; no measurements are done on the reservoir.

The main purpose of this paper is to investigate the robustness of the microscopic reversibility (\ref{microrev}).
Although we assume that the initial state is a product state, we will allow the final state to be arbitrary; 
we will not assume the local system to be macroscopic and we will consider an arbitrary measurement bases.  
We will also assume that 
$\bra{m(T)} \overleftarrow{\hat{\Theta}} \hat{\rho}(T) \hat{\Theta} \ket{m(T)} = \bra{m(T)} \hat{\rho}(T) \ket{m(T)} $; 
the probability to obtain the resulting state $\ket{m(T)}$ is equal to that of the time-reversed state. 
As a result, we will show that the microscopic reversibility does not hold exactly in general; it acquires correction terms.  
The origin of the corrections is the effect that the measurement processes destroy the quantum coherence of the  system that we measure.
%We will also show that the origin of the corrections is the quantum correlation of the system and the reservoir.  
Although the microscopic reversibility is broken in general, 
if we measure the eigenstate of the density matrix of the local system at $t=0$, 
the form of the correction becomes very simple.
In the case of a thermal relaxation process, we numerically confirmed that the correction term is small enough compared to the forward and the reversed transition probabilities in open quantum systems.

This paper is organized as follows:
In Sec.~\ref{isolatedsystem}, we will derive the microscopic reversibility in isolated quantum systems with correction terms under general measurement bases.
As a simple example, we will consider the case of a free particle; 
we will show that a correction term can be very large in this case, and thus we cannot see the microscopic reversibility at all.  
In Sec.~\ref{MRofOQS}, we will derive the microscopic reversibility of open systems with correction terms.  
If the initial and the final states are product states and the eigenstates of the density matrices of the local system are measured, we will show that the correction terms vanish and the microscopic reversibility holds exactly. 
We will also show that, if we disconnect the local system from the reservoir, the correction term then becomes constant.
In Sec.~\ref{PhysicalMeaning}, we analyze the details of the corrections.  
Finally, in Sec.~\ref{SpinChain}, we numerically compute a correction to the microscopic reversibility for a one-dimensional spin chain; 
we regard the first two spins as the local system and the rest as the reservoir.  
The result shows that the correction is relatively small, so that the microscopic reversibility almost holds even when the local system cannot be regarded as macroscopic.

\section{Microscopic reversibility in isolated systems}\label{isolatedsystem}
%We consider the following process:
%The state of an isolated system which is in an arbitrary state at time $t=0$ and
%we measure the state with the projective measurement.
%We denote the measured state as $\ket{n(0)}$. 
%Then we let the system evolve until $t=T$ and measure the state of the system with the projective measurement again.
%We denote that state as $\ket{m(T)}$.

We first describe the microscopic reversibility in isolated quantum systems with the same notations and processes as in Sec.~\ref{Intro}.
The forward and the reversed transition probabilities read \cite{PhysRevLett.100.230404}
\begin{align}
p_{\mathrm{F}}(\ket{n(0)} \rightarrow \ket{m(T)} )
&= \bra{m(T)} \hat{U} \ket{n(0)} \bra{n(0)} \hat{\rho}(0) \ket{n(0)} \bra{n(0)} \hat{U}^{\dagger} \ket{m(T)}, \label{pFiso}\\
p_{\mathrm{R}}(\hat{\Theta}\ket{ m(T)} \rightarrow \hat{\Theta}\ket{ n(0)})
&=\bra{n(0)} \hat{U}^{\dagger} \ket{m(T)}
\bra{m(T)} \overleftarrow{\hat{\Theta}} \hat{\rho}(T) \hat{\Theta} \ket{m(T)} 
\bra{m(T)} \hat{U} \ket{n(0)} \label{pRiso}.
\end{align}
Throughout this paper, we consider the case where 
$\bra{m(T)} \overleftarrow{\hat{\Theta}} \hat{\rho}(T) \hat{\Theta} \ket{m(T)} = \bra{m(T)} \hat{\rho}(T) \ket{m(T)} $. 
This condition is satisfied, for example, in the case where the states or the total Hamiltonian are invariant under the time reversal. 
In Sec.~\ref{ExampleIso}, we treat the former case, the measurement of the position of a particle. 
In Sec.~\ref{SpinChain}, we treat the latter case, the spin system which is invariant under the flip of all the spins.
Then we have the following relation between the forward transition probability and the reversed transition probability: 
\begin{align}
p_{\mathrm{F}}(\ket{n(0)} \rightarrow \ket{m(T)} )
%&= \frac{\bra{n(0)} \hat{\rho}(0) \ket{n(0)}}
%{\bra{m(T)} \hat{\rho}(T) \ket{m(T)}} 
%\bra{n(0)} \hat{U}^{\dagger} \ket{m(T)} 
%\bra{m(T)} \hat{\rho}(T) \ket{m(T)}
%\bra{m(T)} \hat{U} \ket{n(0)} \nonumber\\
&=f_{nm}(T,0) \, p_{\mathrm{R}}(\hat{\Theta} \ket{m(T)} \rightarrow \hat{\Theta} \ket{n(0)}), \label{IsoMR1}
\end{align}
where $f_{nm}(T,0) = \bra{n(0)} \hat{\rho}(0) \ket{n(0)} / \bra{m(T)} \hat{\rho}(T) \ket{m(T)}$.
Let us consider the trivial case where we choose the measurement bases at $t=0$ and $t=T$ as the eigenstates of the density matrices $\hat{\rho}(0)$ and $\hat{\rho}(T)$, respectively.
If we denote $\ket{n^{\prime}(T)}:=\hat{U}\ket{n(0)}$, 
%\textcolor{red}{(MEMO)}$\ket{n^{\prime}(T)}$ is the eigenstate of $\hat{\rho}(T)$, i.e.\,
%\begin{align}
%\hat{\rho}(T) \ket{n^{\prime}(T)}
%:=\hat{\rho}(T) (\hat{U}\ket{n(0)}) = \hat{U} \hat{\rho}(0) \ket{n(0)}
%= p_{n}(0) \hat{U}\ket{n(0)}
%= p_{n}(0) \ket{n^{\prime}(T)},
%\end{align}
we have $\bra{m(T)}\hat{U}\ket{n(0)}=\delta_{n^{\prime}m}$, and therefore
\begin{align}
\bra{n(0)} \hat{\rho}(0) \ket{n(0)}
=\bra{n(0)}\hat{U}^{\dagger} \hat{U} \hat{\rho}(0) \hat{U}^{\dagger} \hat{U}\ket{n(0)}
=\delta_{n^{\prime} m} \bra{m(T)} \hat{\rho}(T) \ket{m(T)}.
\end{align}
Hence the microscopic reversibility trivially holds: 
%for the transition between $\ket{n(0)}$ and $\ket{m(T)}(=\hat{U}\ket{n(0)})$:
\begin{align}
p_{\mathrm{F}}(\ket{n(0)} \rightarrow \ket{m(T)} )
&=p_{\mathrm{R}}(\hat{\Theta}\ket{ m(T)} \rightarrow \hat{\Theta}\ket{ n(0)}). %\delta_{n^{\prime} m}\, 
\end{align}
%which is trivial because the measurements do not disturb the system.

In many cases, however, it is difficult to detect the eigenstates of an arbitrary density matrix or make the system have a density matrix whose eigenstates coincide with the measurement basis that we choose.
In the case where the measurement basis is not the eigenstates of the density matrix, we have
$f_{nm}(T,0)\ne \delta_{n^{\prime} m} $.

In order to observe the effect of the measurement on the microscopic reversibility from a different viewpoint,
let us derive the relation between 
$p_{\mathrm{F}}(\ket{n(0)} \rightarrow \ket{m(T)} )$ and 
$p_{\mathrm{R}}(\hat{\Theta}\ket{ m(T)} \rightarrow \hat{\Theta}\ket{ n(0)})$ 
in the operator-sum representation \cite{nielsen00}.  
The forward and the reversed transition probabilities are, instead of Eqs. (\ref{pFiso}) and (\ref{pRiso}), expressed as
\begin{align}
p_{\mathrm{F}}(\ket{n(0)} \rightarrow \ket{m(T)} )
&= \mathrm{Tr}\, \hat{P}_{m} \hat{U} \hat{P}_{n} \hat{\rho}(0) \hat{P}_{n} \hat{U}^{\dagger} \hat{P}_{m}, \\
p_{\mathrm{R}}(\hat{\Theta}\ket{ m(T)} \rightarrow \hat{\Theta}\ket{ n(0)})
&= \mathrm{Tr}\, \hat{P}_{n} \hat{U}^{\dagger} \hat{P}_{m} \hat{\rho}(T) \hat{P}_{m} \hat{U} \hat{P}_{n},
\end{align}
where
$\hat{P}_{n}=\ket{n(0)}\bra{n(0)}$ and $\hat{P}_{m}=\ket{m(T)}\bra{m(T)}$.
% $\hat{P}_{n}$ and $\hat{P}_{m}$ are the projection operators on $\ket{n}$ at $t=0$ and on $\ket{m}$ at $t=T$ respectively.
Introducing the complementary operator $\hat{Q}_{n}$ and $\hat{Q}_{m}$ of $\hat{P}_{n}$ and $\hat{P}_{m}$, i.e.\,
\begin{align}
\hat{Q}_{k} \equiv \hat{I} - \hat{P}_{k}, \hspace{1cm}
\hat{P}_{k} \hat{Q}_{k}=0, \hspace{1cm} (k=n,m)
\end{align}
we can transform the forward transition probability as follows:
\begin{align}
p_{\mathrm{F}}(\ket{n(0)} \rightarrow \ket{m(T)} )
&= \mathrm{Tr}\, \hat{U}^{\dagger} \hat{P}_{m} \hat{U} 
\hat{P}_{n} \hat{\rho}(0) \hat{P}_{n} \nonumber \\
&= \mathrm{Tr}\, \hat{U}^{\dagger} \hat{P}_{m} \hat{U} 
\big( \hat{\rho}(0) \hat{U}^{\dagger} \hat{U} \hat{\rho}^{-1}(0) \big) 
\hat{P}_{n} \hat{\rho}(0) \hat{P}_{n} \nonumber \\
&= \mathrm{Tr}\, \hat{U}^{\dagger} \hat{P}_{m} \hat{\rho}(T) 
( \hat{P}_{m} + \hat{Q}_{m} )
\hat{U} \hat{\rho}^{-1}(0)
\hat{P}_{n} \hat{\rho}(0) \hat{P}_{n} \nonumber \\
&= \mathrm{Tr}\, \hat{U}^{\dagger} \hat{P}_{m} \hat{\rho}(T) 
\hat{P}_{m} \hat{U} 
( \hat{P}_{n} + \hat{Q}_{n} )
\hat{\rho}^{-1}(0)
\hat{P}_{n} \hat{\rho}(0) \hat{P}_{n} \nonumber\\
&\hspace{30pt} +  \mathrm{Tr}\, \hat{U}^{\dagger} \hat{P}_{m} \hat{\rho}(T) 
\hat{Q}_{m} \hat{U} \hat{\rho}^{-1}(0) \hat{P}_{n} \hat{\rho}(0) \hat{P}_{n} %\sigma_{T} 
\nonumber \\
&= \mathrm{Tr}\, \hat{U}^{\dagger} \hat{P}_{m} \hat{\rho}(T) 
\hat{P}_{m} \hat{U} \hat{P}_{n}
\hat{\rho}^{-1}(0)
\hat{P}_{n} \hat{\rho}(0) \hat{P}_{n} \nonumber\\
&\hspace{30pt} 
+ \mathrm{Tr}\, \hat{U}^{\dagger} \hat{P}_{m} \hat{\rho}(T) 
\hat{Q}_{m} \hat{U} \hat{\rho}^{-1}(0) \hat{P}_{n} \hat{\rho}(0) \hat{P}_{n} %\sigma_{T}
+ \mathrm{Tr}\, \hat{U}^{\dagger} \hat{P}_{m} \hat{\rho}(T) 
\hat{P}_{m} \hat{U} 
\hat{Q}_{n} 
\hat{\rho}^{-1}(0)
\hat{P}_{n} \hat{\rho}(0) \hat{P}_{n}. %\sigma_{0}
\label{IsoMR2}
%&= g_{n} p_{\mathrm{R}}(\hat{\Theta}\ket{ m(T)} \rightarrow \hat{\Theta}\ket{ n(0)}) 
%+ \sigma, \label{IsoMR2}
\end{align}
Therefore, we have
\begin{align}
\frac{ p_{\mathrm{F}}(\ket{n(0)} \rightarrow \ket{m(T)} ) }
{ p_{\mathrm{R}}(\hat{\Theta}\ket{ m(T)} \rightarrow \hat{\Theta}\ket{ n(0)}) }
 = g_{n} + \xi, 
 \label{IsoMR2-1}
\end{align}
where 
\begin{align}
g_{n} &\equiv \bra{n(0)} \hat{\rho}^{-1}(0) \ket{n(0)} \bra{n(0)} \hat{\rho}(0) \ket{n(0)}, \label{iso_g-factor}\\
%\sigma & \equiv 
%\mathrm{Tr}\, \hat{U}^{\dagger} \hat{P}_{m} \hat{\rho}(T) 
%\hat{Q}_{m} \hat{U} \hat{\rho}^{-1}(0) \hat{P}_{n} \hat{\rho}(0) \hat{P}_{n} %\sigma_{T}
%+ \mathrm{Tr}\, \hat{U}^{\dagger} \hat{P}_{m} \hat{\rho}(T) 
%\hat{P}_{m} \hat{U} 
%\hat{Q}_{n} 
%\hat{\rho}^{-1}(0)
%\hat{P}_{n} \hat{\rho}(0) \hat{P}_{n}. %\sigma_{0}
\xi &\equiv 
\frac{
\mathrm{Tr}\, \hat{U}^{\dagger} \hat{P}_{m} \hat{\rho}(T) 
\hat{Q}_{m} \hat{U} \hat{\rho}^{-1}(0) \hat{P}_{n} \hat{\rho}(0) \hat{P}_{n} %\sigma_{T}
+ \mathrm{Tr}\, \hat{U}^{\dagger} \hat{P}_{m} \hat{\rho}(T) 
\hat{P}_{m} \hat{U} 
\hat{Q}_{n} 
\hat{\rho}^{-1}(0)
\hat{P}_{n} \hat{\rho}(0) \hat{P}_{n} }
{ p_{\mathrm{R}}(\hat{\Theta}\ket{ m(T)} \rightarrow \hat{\Theta}\ket{ n(0)}) }. %\sigma_{0}
\end{align}
We assumed that $\hat{\rho}(0)$ is the state where its inverse exists.
In order to see the structure of the correction term $\xi$ more explicitly, 
let us divide it in the following way:
\begin{align}
\xi 
&= p_{\mathrm{R}}^{-1} \, \mathrm{Tr} \, \hat{U}^{\dagger}
 \hat{P}_{m} \hat{\rho}(T) 
 ( \hat{Q}_{m} \hat{U} + \hat{P}_{m} \hat{U} \hat{Q}_{n} )
 \hat{\rho}^{-1}(0) \hat{P}_{n} 
  \hat{\rho}(0) \hat{P}_{n} \nonumber\\
&= p_{\mathrm{R}}^{-1} \, \mathrm{Tr} \, \hat{U}^{\dagger}
 \hat{P}_{m} \hat{\rho}(T) 
 [ \hat{Q}_{m} \hat{U} (\hat{P}_{n} + \hat{Q}_{n} ) + \hat{P}_{m} \hat{U} \hat{Q}_{n} ]
 \hat{\rho}^{-1}(0) \hat{P}_{n} 
  \hat{\rho}(0) \hat{P}_{n} \nonumber\\
&= p_{\mathrm{R}}^{-1} \, \mathrm{Tr} \, \hat{U}^{\dagger}
 \hat{P}_{m} \hat{\rho}(T) 
 ( \hat{Q}_{m} \hat{U} \hat{P}_{n}  + \hat{U} \hat{Q}_{n} )
 \hat{\rho}^{-1}(0) \hat{P}_{n} 
  \hat{\rho}(0) \hat{P}_{n} \nonumber\\
&= p_{\mathrm{R}}^{-1} \, 
 g_{n}\, \mathrm{Tr} \, \hat{U}^{\dagger}
 \hat{P}_{m} \hat{\rho}(T) 
  \hat{Q}_{m} \hat{U} \hat{P}_{n} 
  + p_{\mathrm{R}}^{-1} \,  \mathrm{Tr} \, \hat{U}^{\dagger}
 \hat{P}_{m} \hat{\rho}(T) 
  \hat{U} \hat{Q}_{n} 
 \hat{\rho}^{-1}(0) \hat{P}_{n} 
  \hat{\rho}(0) \hat{P}_{n} \nonumber\\
&=: \xi_{\alpha} + \xi_{\beta}. 
\label{xi_partition}
\end{align}

As a special case, let us consider the situation where we measure the density matrix at $t=0$ with the eigenstate basis and obtain the result $\ket{n(0)}$, i.e.\,, 
$\hat{\rho}(0) \hat{P}_{n} = p_{n} \hat{P}_{n}$ and $\hat{\rho}^{-1}(0) \hat{P}_{n} = p^{-1}_{n} \hat{P}_{n}$ where $ p_{n}$ is an eigenvalue of the density matrix $\hat{\rho}(0)$ for $\ket{n(0)}$.
%, e.g., the initial state is the canonical equilibrium state and its energy is measured. 
Then we have $g_{n}=p_{n} p^{-1}_{n}=1$ and 
$\xi_{\beta} = 0$, but $\xi_{\alpha} \ne 0 \in \mathcal{R}$:
\begin{align}
&\frac{ p_{\mathrm{F}}(\ket{n(0)} \rightarrow \ket{m(T)} ) }
{ p_{\mathrm{R}}(\hat{\Theta}\ket{ m(T)} \rightarrow \hat{\Theta}\ket{ n(0)}) } =
1 + \xi_{\alpha}, \label{specialpF}
\end{align}
where 
\begin{align}
&\xi_{\alpha} =  p_{\mathrm{R}}^{-1} \, \mathrm{Tr}\, \hat{U}^{\dagger} \hat{P}_{m} \hat{\rho}(T) 
\hat{Q}_{m} \hat{U} \hat{P}_{n}.  \label{specialsigma}
\end{align}
Using (\ref{IsoMR1}), we can relate the correction term $\xi_{\alpha}$ to $f_{nm}(T,0)$ as 
\begin{align}
f_{nm}(T,0) = 1 + \xi_{\alpha}.\label{f_vs_sigma}
%\sigma_{T} &= (1-f_{nm}(T,0)^{-1}) p_{\mathrm{F}}(\ket{n(0)} \rightarrow \ket{m(T)} ) \nonumber\\
%&= (f_{nm}(T,0) - 1) p_{\mathrm{R}}(\hat{\Theta}\ket{ m(T)} \rightarrow \hat{\Theta}\ket{ n(0)}).\label{f_vs_sigma}
\end{align}
%Therefore, if $f_{nm}(T,0)>1$, i.e.\,$\bra{n(0)} \hat{\rho}(0) \ket{n(0)} > \bra{m(T)} \hat{\rho}(T) \ket{m(T)}$,
%the forward probability is greater by $\sigma_{T}$ in (\ref{f_vs_sigma}) than the reversed probability, and vice versa.
When we measure the eigenstate of the density matrix at $t=T$, on the other hand, we have $\xi_{\alpha}=0$. 
As we will explain in detail in Sec.~\ref{PhysicalMeaning} for open systems, the corrections $g_{n}$, $\xi_{\alpha}$, and $\xi_{\beta}$ depend on the choices of the measurement bases and, in the case of an open system, on the state of the total system as well.
We can use $g_{n}, \xi_{\alpha}$ and $\xi_{\alpha}$ as the measures of the irreversibility caused by the measurement processes.
The analysis of (\ref{IsoMR2-1}) is more advantageous than (\ref{IsoMR1}) in the case of open quantum systems as we will see in Sec.\ref{MRofOQS}.

\subsection{Example: One-dimensional free particle} \label{ExampleIso}
\begin{figure}[th]
%\begin{minipage}{0.5\hsize}
\begin{center}
\includegraphics[width=70mm,clip]{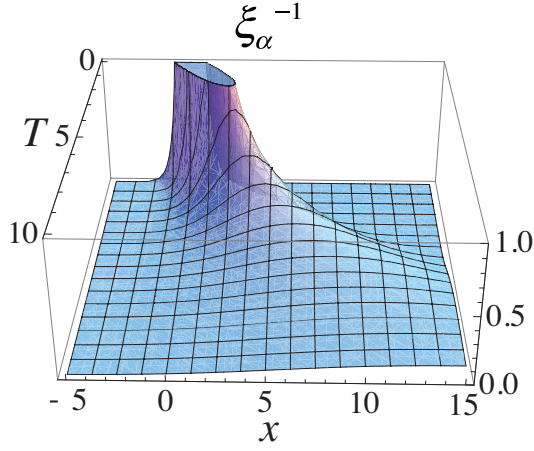}
%(a)
\end{center}
%\end{minipage}
%\begin{minipage}{0.5\hsize}
%\begin{center}
%\includegraphics[width=60mm,clip]{PF_Sigma_iso.eps}
%(b)
%\end{center}
%\end{minipage}
\caption{
(Color online) The $T$ dependence of $\xi_{\alpha}$ 
with $p^{(1)}_{0}=1$, $p^{(2)}_{0}=2$, $M=1$, and $a=1.2$.
}
\label{isoMRfree}
\end{figure}

A free particle in one dimension is the simplest illustrative example.
We assume that the initial state is given by
\begin{align}
\hat{\rho}(0) = \frac{1}{2} \ket{\phi_{1}(0)}\bra{\phi_{1}(0)}  + \frac{1}{2} \ket{\phi_{2}(0)}\bra{\phi_{2}(0)},
\end{align} 
where $\ket{\phi_{k}(0)}$ $(k=1,2)$ are the Gaussian wave packets in the position representation:
\begin{align}
\braket{x}{\phi_{k}(0)} = \frac{1}{\left( a^{2}\pi \right)^{1/4}} \exp\left( ip^{(k)}_{0} x - \frac{x^{2}}{2a^{2}} \right),
\end{align}
where $x$ and $p^{(k)}_{0}$ are the position and the momentum of the particle and the parameter $a$ determines the width of the wave packet.
Denoting $M$ as the mass of the particle, we can write the state at $t=T$ as 
\begin{align}
\hat{\rho}(T) &= \frac{1}{2} \ket{\phi_{1}(T)}\bra{\phi_{1}(T)}  + \frac{1}{2} \ket{\phi_{2}(T)}\bra{\phi_{2}(T)}, \\
\braket{x}{\phi_{k}(T)} &= \left[ \sqrt{\pi} \left(a + \frac{iT}{M a} \right) \right]^{-\frac{1}{2}} 
\exp\biggl[ \left(a + \frac{iT}{M a} \right)^{-1} 
a \left( -\frac{x^{2}}{2a^{2}}+ip^{(k)}_{0}x - \frac{i (p^{(k)}_{0})^{2}T}{2M} \right)
\biggr].
\end{align} 
We measure the initial state with the basis which contains $\ket{\phi_{1}(0)}$ and $\ket{\phi_{2}(0)}$ 
and measure the position $x$ at $t=T$; 
we have $g_{n} = 1$ and $\xi_{\beta}= 0$.
For the transition from $\phi_{1}(0)$ to $x(T)$, we have
\begin{align}
\bra{\phi_{1}(0)} \hat{\rho}(0) \ket{\phi_{1}(0)} &= \frac{1}{2}, \nonumber\\
\bra{x(T)} \hat{\rho}(T) \ket{x(T)} 
&= \frac{1}{2} |\braket{x(T)}{\phi_{1}(T)} |^{2} +
\frac{1}{2} |\braket{x(T)}{\phi_{2}(T)} |^{2},
\nonumber\\
p_{\mathrm{F}}(\ket{n(0)} \rightarrow \ket{m(T)} )
&= |\bra{x(T)}\hat{U}\ket{\phi_{1}(0)}|^{2} \bra{\phi_{1}(0)} \hat{\rho}(0) \ket{\phi_{1}(0)}
= \frac{1}{2} |\braket{x(T)}{\phi_{1}(T)}|^{2},
\end{align}
which, according to (\ref{f_vs_sigma}), gives 
\begin{align}
\xi_{\alpha} = ( |\braket{x(T)}{\phi_{1}(T)} |^{2} + |\braket{x(T)}{\phi_{2}(T)} |^{2} )^{-1} - 1.
\end{align}
The $T$ dependences of $\xi_{\alpha}$ is plotted in Fig.~\ref{isoMRfree}. 
Here, we set $p^{(1)}_{0}=1$, $p^{(2)}_{0}=2$, $M=1$, and $a=1.2$.
The inverse of the correction term $\xi_{\alpha}^{-1}$ vanishes as $x$ and $T$ increase, 
which means that the correction grows much larger than unity; 
we cannot see the microscopic reversibility at all in this example.

For the isolated quantum systems, we confirmed that the microscopic reversibility is not a general relation. 
Because there is nothing like thermalization, the deviation from the microscopic reversibility seems to depend sensitively on the choice of the system and the protocol; the microscopic reversibility is not a proper relation to characterize the dynamics of the isolated quantum systems.
For the open quantum systems, nevertheless, we expect the microscopic reversibility is indeed a proper relation due to the effect of the thermalization; we will show in Sec.~\ref{SpinChain} that our expectation seems to be correct.

\section{Microscopic reversibility in open quantum systems}\label{MRofOQS}
We next extend the discussion in Sec.~\ref{isolatedsystem} to the case of open quantum systems. 
Let us consider the microscopic reversibility of the local quantum system which is thrown into a reservoir at $t=0$.
We allow the local system and the reservoir to be externally controlled by time-dependent parameters $\lambda_{\mathrm{s}}(t)$ and $\lambda_{\mathrm{r}}(t)$.
The total Hamiltonian reads
\begin{align}
\hat{H}_{\mathrm{tot}}(t) = \hat{H}_{\mathrm{s}}(\lambda_{\mathrm{s}}(t)) + \hat{H}_{\mathrm{r}}(\lambda_{\mathrm{r}}(t)) + \hat{H}_{\mathrm{c}}\theta(t),
\end{align}
where $\hat{H}_{\mathrm{s}}(\lambda_{\mathrm{s}}(t))$, $\hat{H}_{\mathrm{r}}(\lambda_{\mathrm{r}}(t))$, and $\hat{H}_{\mathrm{c}}$ are the Hamiltonians of the local system, the reservoir, and the coupling between the local system and the reservoir, respectively. 
The function $\theta(t)$ is a step function.
%The Hamiltonian of the local system $\hat{H}_{\mathrm{s}}(t)$ is time-dependent because it may be driven by some external field.
Since we consider the situation where the coupling is turned on at $t=0$, the initial state is given as a product state
\begin{align}
\hat{\rho}_{\mathrm{tot}}(0)= \hat{\rho}_{\mathrm{s}}(0) \otimes \hat{\rho}_{\mathrm{r}}(0). \label{ProductState}
\end{align}
As we mentioned in Sec.~\ref{Intro},
we consider the forward and the reversed transition probabilities of the local system under the constraint that 
we measure the local system only.
The transition probability of the forward process that the state of the local system evolves from $\ket{n(0)}$ to $\ket{m(T)}$ reads
\begin{align}
p_{\mathrm{F}}(\ket{n(0)} \rightarrow \ket{m(T)} )
&=\mathrm{Tr}_{\mathrm{r}} \bra{m(T)} \hat{U} \ket{n(0)} \bra{n(0)} \hat{\rho}_{\mathrm{tot}}(0) \ket{n(0)} \bra{n(0)} \hat{U}^{\dagger} \ket{m(T)} 
\end{align}
and the reversed transition probability that the state of the local system evolves from $\hat{\Theta} \ket{m(T)}$ to $\hat{\Theta} \ket{n(0)}$ reads
\begin{align}
p_{\mathrm{R}}(\hat{\Theta}\ket{m(T)} \rightarrow \hat{\Theta}\ket{n(0)} )
&=\mathrm{Tr}_{\mathrm{r}} \bra{n(0)} \hat{U}^{\dagger} \ket{m(T)} \bra{m(T)} \hat{\rho}_{\mathrm{tot}}(T) \ket{m(T)} \bra{m(T)} \hat{U} \ket{n(0)}, 
\end{align}
where $\bra{m(T)} \hat{U} \ket{n(0)}$, 
$\bra{n(0)} \hat{\rho}_{\mathrm{tot}}(0) \ket{n(0)}$, 
and $\bra{n(0)} \hat{U}^{\dagger} \ket{m(T)}$ are the operators on the reservoir,
whereas $\mathrm{Tr}_{\mathrm{r}}$ is the trace with respect to the degrees of freedom of the reservoir.
As in the case of isolated systems, we again assumed that 
$\bra{m(T)} \hat{\Theta} \hat{\rho}_{\mathrm{tot}}(T) \hat{\Theta} \ket{m(T)} = \bra{m(T)} \hat{\rho}_{\mathrm{tot}}(T) \ket{m(T)}$. 
Note that $p_{\mathrm{F}}(\ket{n(0)} \rightarrow \ket{m(T)} )$ is indeed a 
%non-negative 
real number since $\hat{\rho}_{\mathrm{tot}}(0)$ is Hermitian:
\begin{align}
&p_{\mathrm{F}}(\ket{n(0)} \rightarrow \ket{m(T)} )^{\ast} \nonumber\\
&=\sum_{x,x^{\prime},y} \biggl( \bra{m(T),y} \hat{U} \ket{n(0),x} \bra{n(0),x} \hat{\rho}_{\mathrm{tot}}(0) \ket{n(0),x^{\prime}} \bra{n(0),x^{\prime}} \hat{U}^{\dagger} \ket{m(T),y} \biggr)^{\ast} \nonumber\\
&=\sum_{x,x^{\prime},y} \bra{m(T),y} \hat{U} \ket{n(0),x^{\prime}} \bra{n(0),x^{\prime}} \hat{\rho}_{\mathrm{tot}}(0) \ket{n(0),x} \bra{n(0),x} \hat{U}^{\dagger} \ket{m(T),y} \nonumber\\
&=p_{\mathrm{F}}(\ket{n(0)} \rightarrow \ket{m(T)} ),
\end{align}
where $x,\, x^{\prime}$, and $y$ are the states of the reservoir.

As we did for isolated systems, 
let us introduce the projection operators $\hat{P}_{n}$ to project on the state $\ket{n(0)}$ of the local system at time $t=0$ 
and $\hat{Q}_{n}$ to project on the complementary space of $\hat{P}_{n}$, i.e.\,, 
\begin{align}
&\hat{P}_{n} \equiv \sum_{x(0)} \ket{n(0), x(0)} \bra{n(0), x(0)},\hspace{10pt}
\hat{Q}_{n} \equiv \hat{I}- \hat{P}_{n}. 
\end{align}
Similarly, we define $\hat{P}_{m}$ to project on the state $\ket{m(T)}$ of the local system at time $t=T$ 
and $\hat{Q}_{m}$ to project on the complementary space of $\hat{P}_{m}$, i.e.\,, 
\begin{align}
&\hat{P}_{m} \equiv \sum_{x(T)} \ket{m(T), x(T)} \bra{m(T), x(T)},\hspace{10pt}
\hat{Q}_{m} \equiv \hat{I}- \hat{P}_{m}. 
\end{align}

Now, we rewrite the forward probability as follows:
\begin{align}
p_{\mathrm{F}}(\ket{n(0)} \rightarrow \ket{m(T)} ) 
&=\mathrm{Tr}\, \hat{U}^{\dagger} P_{m} \hat{U} \hat{P}_{n} \hat{\rho}_{\mathrm{tot}}(0) \hat{P}_{n} \nonumber\\
&=\mathrm{Tr}\, \hat{U}^{\dagger} P_{m} 
\hat{U} \hat{\rho}_{\mathrm{tot}}(0) \hat{U}^{\dagger} \hat{U} \hat{\rho}^{-1}_{\mathrm{tot}}(0) 
\hat{P}_{n} \hat{\rho}_{\mathrm{tot}}(0) \hat{P}_{n} \nonumber\\
&=\mathrm{Tr}\, \hat{U}^{\dagger} P_{m} 
\hat{\rho}_{\mathrm{tot}}(T) \hat{U} \hat{\rho}^{-1}_{\mathrm{tot}}(0) 
\hat{P}_{n} \hat{\rho}_{\mathrm{tot}}(0) \hat{P}_{n} \nonumber\\ 
&=\mathrm{Tr}\, \hat{U}^{\dagger} P_{m} 
\hat{\rho}_{\mathrm{tot}}(T) (\hat{P}_{m} + \hat{Q}_{m}) \hat{U} \hat{\rho}^{-1}_{\mathrm{tot}}(0) 
\hat{P}_{n} \hat{\rho}_{\mathrm{tot}}(0) \hat{P}_{n} \nonumber\\ 
&=\mathrm{Tr}\, \hat{U}^{\dagger} P_{m} 
\hat{\rho}_{\mathrm{tot}}(T) \hat{P}_{m} \hat{U} \hat{\rho}^{-1}_{\mathrm{tot}}(0) 
\hat{P}_{n} \hat{\rho}_{\mathrm{tot}}(0) \hat{P}_{n} \nonumber\\
&\hspace{30pt} + \mathrm{Tr}\, \hat{U}^{\dagger} P_{m} 
\hat{\rho}_{\mathrm{tot}}(T) \hat{Q}_{m} \hat{U} \hat{\rho}^{-1}_{\mathrm{tot}}(0) 
\hat{P}_{n} \hat{\rho}_{\mathrm{tot}}(0) \hat{P}_{n} \nonumber\\ 
&=\mathrm{Tr}\, \hat{U}^{\dagger} P_{m} 
\hat{\rho}_{\mathrm{tot}}(T) \hat{P}_{m} \hat{U} (\hat{P}_{n} + \hat{Q}_{n}) \hat{\rho}^{-1}_{\mathrm{tot}}(0) 
\hat{P}_{n} \hat{\rho}_{\mathrm{tot}}(0) \hat{P}_{n} \nonumber\\
&\hspace{30pt} + \mathrm{Tr}\, \hat{U}^{\dagger} P_{m} 
\hat{\rho}_{\mathrm{tot}}(T) \hat{Q}_{m} \hat{U} \hat{\rho}^{-1}_{\mathrm{tot}}(0) 
\hat{P}_{n} \hat{\rho}_{\mathrm{tot}}(0) \hat{P}_{n} \nonumber\\ 
&=\mathrm{Tr}\, \hat{U}^{\dagger} P_{m} 
\hat{\rho}_{\mathrm{tot}}(T) \hat{P}_{m} \hat{U} \hat{P}_{n} \hat{\rho}^{-1}_{\mathrm{tot}}(0) 
\hat{P}_{n} \hat{\rho}_{\mathrm{tot}}(0) \hat{P}_{n} \nonumber\\
&\hspace{30pt} +\mathrm{Tr}\, \hat{U}^{\dagger} P_{m} 
\hat{\rho}_{\mathrm{tot}}(T) \hat{P}_{m} \hat{U} \hat{Q}_{n} \hat{\rho}^{-1}_{\mathrm{tot}}(0) 
\hat{P}_{n} \hat{\rho}_{\mathrm{tot}}(0) \hat{P}_{n} \nonumber\\
&\hspace{30pt} + \mathrm{Tr}\, \hat{U}^{\dagger} P_{m} 
\hat{\rho}_{\mathrm{tot}}(T) \hat{Q}_{m} \hat{U} \hat{\rho}^{-1}_{\mathrm{tot}}(0) 
\hat{P}_{n} \hat{\rho}_{\mathrm{tot}}(0) \hat{P}_{n}. 
\label{PF1}
\end{align}
We assume that the initial state is the product state (\ref{ProductState}), 
but the measurement bases are not necessarily the eigenstates of the density matrix of the local system,
and thus the first term of Eq.~(\ref{PF1}) reads
\begin{align}
\mathrm{Tr}\, \hat{U}^{\dagger} P_{m} 
\hat{\rho}_{\mathrm{tot}}(T) \hat{P}_{m} \hat{U} \hat{P}_{n} \hat{\rho}^{-1}_{\mathrm{tot}}(0) 
\hat{P}_{n} \hat{\rho}_{\mathrm{tot}}(0) \hat{P}_{n} 
&= g_{n}\, p_{\mathrm{R}}(\hat{\Theta} \ket{m(T)} \rightarrow \hat{\Theta} \ket{n(0)}), 
\end{align}
where
\begin{align}
g_{n} \equiv \bra{n(0)} \hat{\rho}^{-1}_{\mathrm{s}}(0) \ket{n(0)}  \bra{n(0)} \hat{\rho}_{\mathrm{s}}(0) \ket{n(0)}.
\label{g-factor}
\end{align}
We do exactly the same transform as Eq.~(\ref{xi_partition}) for the second and third term except that the density matrix is $\hat{\rho}_{\mathrm{tot}}$ and the projection operators are for the local system only. 
Then we arrive at one of the major results of the present paper: 
\begin{align}
\frac{ p_{\mathrm{F}}(\ket{n(0)} \rightarrow \ket{m(T)} ) }
{ p_{\mathrm{R}}(\hat{\Theta}\ket{ m(T)} \rightarrow \hat{\Theta}\ket{ n(0)}) }
 = g_{n} + \xi_{\alpha} + \xi_{\beta}, 
 \label{openMR}
\end{align}
where 
\begin{align}
\xi_{\alpha} &\equiv p_{\mathrm{R}}^{-1} \, 
 g_{n}\, \mathrm{Tr} \, \hat{U}^{\dagger}
 \hat{P}_{m} \hat{\rho}_{\mathrm{tot}}(T) 
  \hat{Q}_{m} \hat{U} \hat{P}_{n}, \nonumber \\
\xi_{\beta} &\equiv  p_{\mathrm{R}}^{-1} \,  \mathrm{Tr} \, \hat{U}^{\dagger}
 \hat{P}_{m} \hat{\rho}_{\mathrm{tot}}(T) 
  \hat{U} \hat{Q}_{n} 
 \hat{\rho}^{-1}_{\mathrm{tot}}(0) \hat{P}_{n} 
  \hat{\rho}_{\mathrm{tot}}(0) \hat{P}_{n}.
\end{align}
We do not assume that the final state to be a product state.
We cannot analyze the microscopic reversibility in the same way as in (\ref{IsoMR2-1}) in the case of open quantum systems, because the elements $\bra{n(0)} \hat{\rho}_{\mathrm{tot}}(0) \ket{n(0)}$ and $\bra{m(T)} \hat{\rho}_{\mathrm{tot}}(T) \ket{m(T)}$ are not c-numbers.
Equation (\ref{openMR}) reduces to (\ref{IsoMR2-1}) by eliminating the degrees of freedom of the reservoir.
We will analyze the properties and the meanings of the corrections in Eq.~(\ref{openMR}) in Sec.~\ref{PhysicalMeaning}.

\subsection{A case where the initial and the final states are product states}\label{NoCorrections}
If the density matrix of the local system $\hat{\rho}_{\mathrm{s}}(0)$ in (\ref{ProductState}) is measured at $t=0$ and the result is an eigenstate $\ket{n(0)}$,
we have
$\hat{\rho}_{\mathrm{s}}(0) \hat{P}_{n} = p_{n} \hat{P}_{n}$, 
where $p_{n}$ is the eigenvalue of $\hat{\rho}_{\mathrm{s}}(0)$.
Therefore, just as in the case of isolated systems, 
$g_{n}=p^{-1}_{n} p_{n}=1$ and 
$\xi_{\beta} = 0$.
If the final state is also a product state and the density matrix of the local system is measured at $t=T$ with the result of an eigenstate $\ket{m(T)}$, 
we have 
$\hat{\rho}_{\mathrm{s}}(T) \hat{P}_{m} = p_{m} \hat{P}_{m}$, 
where $p_{m}$ is the eigenvalue of $\hat{\rho}_{\mathrm{s}}(T)$; 
we have 
$\xi_{\alpha}=0$ in this case.
When both of these conditions are satisfied, the microscopic reversibility holds exactly:
\begin{align}
p_{\mathrm{F}}(\ket{n(0)} \rightarrow \ket{m(T))}
=  p_{\mathrm{R}}(\hat{\Theta} \ket{m(T)} \rightarrow \hat{\Theta} \ket{n(0)}).\label{MR2}
\end{align}
This is another major results of the present paper.  
This means that no matter how strong the local system is connected to the reservoir 
during the period between the measurements,
the microscopic reversibility holds as long as the above conditions are satisfied.
Note that this is a sufficient condition; we are not yet sure what is the necessary condition to make the correction terms vanish.

\subsection{A case where the local system is disconnected from the reservoir}\label{Disconnect}
Again, we consider the case where we measure the density matrix of the local system at $t=0$ with the result of an eigenstate $\ket{n(0)}$.
Now, let us consider the process where we gradually disconnect the local system from the reservoir
and measure the energy of the local system at $t=T$ with the result of an energy eigenstate $\ket{m(T)}$.
Here we do not assume the form of the density matrix at $t=T$.

If we split the time evolution at the time $\tau_{\mathrm{iso}}$, at which we can regard that the local system is almost isolated from the reservoir, i.e.\,
\begin{align}
\hat{U}(T) = \hat{U}(T-\tau_{\mathrm{iso}}) \, \hat{U}(\tau_{\mathrm{iso}})
=\left( \exp\left( -i \hat{H}_{\mathrm{s}} (T-\tau_{\mathrm{iso}})\right) \otimes
\exp\left( -i \hat{H}_{\mathrm{r}} (T-\tau_{\mathrm{iso}})\right) \right)
\, \hat{U}(\tau_{\mathrm{iso}}), 
\end{align}
then we have
\begin{align}
p_{\mathrm{F}}(\ket{n(0)} \rightarrow \ket{m(T)} )
&=\mathrm{Tr}_{\mathrm{r}} \bra{m(T)} \hat{U}(\tau_{\mathrm{iso}}) \ket{n(0)} \bra{n(0)} \hat{\rho}_{\mathrm{tot}}(0) \ket{n(0)} \bra{n(0)} \hat{U}^{\dagger}(\tau_{\mathrm{iso}}) \ket{m(T)}, \nonumber\\
p_{\mathrm{R}}(\Theta \ket{m(T)} \rightarrow \Theta \ket{n(0)})
&=\mathrm{Tr}_{\mathrm{r}} 
\bra{n(0)} \hat{U}^{\dagger}(\tau_{\mathrm{iso}}) 
\ket{m(T)} \bra{m(T)} 
\hat{U}(\tau_{\mathrm{iso}}) \hat{\rho}_{\mathrm{tot}}(0) \hat{U}^{\dagger}(\tau_{\mathrm{iso}}) 
\ket{m(T)} \bra{m(T)} 
\hat{U}(\tau_{\mathrm{iso}}) \ket{n(0)},
\end{align}
which are independent of the state at $t=T$.
Therefore, the transition probabilities and the correction term $\xi_{\alpha}$ become constant after the local system is completely disconnected.

Compared to the case of the isolated quantum systems, the behavior of the microscopic reversibility in open quantum systems is more nontrivial depending on the protocol that we choose. 
%In the following Section, we will rewrite the correction terms $\sigma_{0}$ and $\sigma_{T}$ in order to digest more about those nontrivial behavior. 

\section{Details of the corrections}\label{PhysicalMeaning}
We only consider the case of open quantum systems because the consequences for isolated systems follow by eliminating the degrees of freedom of the reservoir.

\subsection{correction factor $g_{n}$}
Let us analyze the properties of the correction factor $g_{n}$.  
Introducing a unitary matrix $\hat{V}$ that transforms the density matrix of the local system $\hat{\rho}_{\mathrm{s}}(0)$ in the present measurement basis to its diagonal form, i.e.
\begin{align}
\hat{\rho}_{\mathrm{s}}(0) = \hat{V} \hat{\rho}_{D}(0) \hat{V}^{\dagger},  \hspace{1cm} 
\hat{\rho}_{D}(0) = 
\mathrm{diag} (p_{1}, p_{2},\cdots, p_{N}), 
%\begin{pmatrix}
%p_{1} &0&\cdots&0\\
%0&p_{2}&\cdots&0\\
%\vdots&\vdots&\ddots&\\
%0&0&&p_{N}
%\end{pmatrix}, 
\end{align}
we can write the general form of $g_{n}$ in (\ref{g-factor}) as
\begin{align}
g_{n}=\sum_{i,j=1}^{N} \frac{p_{i}}{p_{j}} \left| v_{ni} \right|^{2} \left| v_{nj} \right|^{2}, \label{ExplicitForm}
\end{align}
where $v_{ij}$ is the $(i,j)$ element of $\hat{V}$. 
We have $g_{n}=1$ if we choose the state $\ket{n(0)}$ as an eigenstate of the density matrix of the local system.
Since $\hat{\rho}_{\mathrm{s}}(0)$ is a positive operator, $g_{n}$ is positive, which is also obvious from (\ref{ExplicitForm}).
The factor $g_{n}$ can be large if some states have a relatively small probability $p_{j}$ in the diagonalizing basis.
Note, however, that $g_{n}$ is finite; to have $g_{n}=\infty$, it would require that $p_{j}=0$ for the $j$th state, 
in which case the inverse of $\hat{\rho}_{\mathrm{s}}(0)$ would not exist.

\subsection{correction terms $\xi_{\alpha}$ and $\xi_{\beta}$}
The correction terms from the microscopic reversibility must be quantities related to the disturbance due to the measurement process.
%We now rewrite the correction terms so that their meanings are more explicit.
Here, we consider the quantities $\sigma_{\alpha} \equiv \xi_{\alpha} \, p_{\mathrm{R}}$ and $\sigma_{\beta} \equiv \xi_{\beta} \, p_{\mathrm{R}}$.  
We recast these correction terms into the forms
\begin{align}
\sigma_{\alpha} 
&= g_{n}\, \mathrm{Tr} \, \hat{P}_{n} \hat{U}^{\dagger} \hat{P}_{m} \hat{\rho}_{\mathrm{tot}}(T) (1- \hat{P}_{m}) \hat{U} \nonumber \\
&= g_{n}\, \mathrm{Tr} \, \hat{P}_{n} \hat{U}^{\dagger} \hat{P}_{m} \hat{\rho}_{\mathrm{tot}}(T) \hat{U} 
- g_{n}\, \mathrm{Tr} \, \hat{P}_{n} \hat{U}^{\dagger} \hat{P}_{m} \hat{\rho}_{\mathrm{tot}}(T) \hat{P}_{m} \hat{U} \nonumber\\
&= g_{n}\, \mathrm{Tr} \, \hat{P}_{n} \hat{U}^{\dagger} \hat{P}_{m} \hat{\rho}_{\mathrm{tot}}(T) \hat{U}
 - g_{n}\, p_{\mathrm{R}}(\hat{\Theta} \ket{m(T)} \rightarrow \hat{\Theta} \ket{n(0)}), \label{sigma_alpha}
\end{align}
and
\begin{align}
\sigma_{\beta} &= \mathrm{Tr} \, 
\hat{P}_{n} \hat{U}^{\dagger} \hat{P}_{m} \hat{\rho}_{\mathrm{tot}}(T) \hat{U}
(1- \hat{P}_{n} )\hat{\rho}^{-1}_{\mathrm{tot}}(0) \hat{P}_{n} \hat{\rho}_{\mathrm{tot}}(0)\nonumber\\
 &= \mathrm{Tr} \, 
\hat{P}_{n} \hat{U}^{\dagger} \hat{P}_{m} \hat{\rho}_{\mathrm{tot}}(T) \hat{U}
\hat{\rho}^{-1}_{\mathrm{tot}}(0) \hat{P}_{n} \hat{\rho}_{\mathrm{tot}}(0)
- g_{n}\, \mathrm{Tr} \, \hat{P}_{n} \hat{U}^{\dagger} \hat{P}_{m} \hat{\rho}_{\mathrm{tot}}(T) \hat{U} \nonumber\\
 &= \mathrm{Tr} \, 
\hat{P}_{n} \hat{U}^{\dagger} \hat{P}_{m} \hat{U}
\hat{P}_{n} \hat{\rho}_{\mathrm{tot}}(0)
- g_{n}\, \mathrm{Tr} \, \hat{P}_{n} \hat{U}^{\dagger} \hat{P}_{m} \hat{\rho}_{\mathrm{tot}}(T) \hat{U} \nonumber\\
&=p_{\mathrm{F}}(\ket{n(0)} \rightarrow \ket{m(T)}) 
-  g_{n}\, \mathrm{Tr} \, \hat{P}_{n} \hat{U}^{\dagger} \hat{P}_{m} \hat{\rho}_{\mathrm{tot}}(T) \hat{U}. \label{sigma_beta}
\end{align}

%Taking the sum over the final states, we have
%\begin{align}
%\sum_{m} \sigma_{\alpha} 
%&= \sum_{m}\left[
%g_{n}\, \mathrm{Tr} \, \hat{P}_{n} \hat{U}^{\dagger} \hat{P}_{m} \hat{\rho}_{\mathrm{tot}}(T) \hat{U}
% - g_{n}\, p_{\mathrm{R}}(\hat{\Theta} \ket{m(T)} \rightarrow \hat{\Theta} \ket{n(0)})
%\right]\nonumber \\
%&= g_{n}\, \mathrm{Tr} \, \hat{P}_{n} \hat{U}^{\dagger} \left( \sum_{m} \hat{P}_{m} \right) \hat{U} \hat{\rho}_{\mathrm{tot}}(0) \hat{U}^{\dagger} \hat{U}
% - g_{n} \sum_{m} p_{\mathrm{R}}(\hat{\Theta} \ket{m(T)} \rightarrow \hat{\Theta} \ket{n(0)}) \nonumber \\
%&=g_{n} \left\{ \mathrm{Tr} P_{n}\hat{\rho}_{\mathrm{tot}}(0)  - \sum_{m}\,p_{\mathrm{R}}(\hat{\Theta} \ket{m(T)} \rightarrow \hat{\Theta} \ket{n(0)}) \right\},\\
%\sum_{m} \sigma_{\beta} 
%&=\sum_{m}\, \left[ p_{\mathrm{F}}(\ket{n(0)} \rightarrow \ket{m(T)}) 
%-  g_{n}\, \mathrm{Tr} \, \hat{P}_{n} \hat{U}^{\dagger} \hat{P}_{m} \hat{\rho}_{\mathrm{tot}}(T) \hat{U}
%\right] \nonumber\\
%&= \mathrm{Tr} P_{n}\hat{\rho}_{\mathrm{tot}}(0) \, (1 - g_{n}).
%\end{align}
%This indicates that the correction terms are the quantities depending on the choice of the measurement basis.
%Taking the sum over the final states and consider the case where 
%the measurement of the initial state is of the eigenstate of the local system, i.e.\,, $g_{n}=1$,
Taking the sum over the final states, 
we have from the third equality of (\ref{sigma_alpha}) and 
the second equality of (\ref{sigma_beta}),
\begin{align}
\sum_{m} \sigma_{\alpha} 
&=\sum_{m} \bigl( g_{n} \mathrm{Tr} \, \hat{P}_{n} \hat{U}^{\dagger} \hat{P}_{m} \hat{\rho}_{\mathrm{tot}}(T) \hat{U} 
- g_{n} \mathrm{Tr} \, \hat{P}_{n} \hat{U}^{\dagger} \hat{P}_{m} \hat{\rho}_{\mathrm{tot}}(T) \hat{P}_{m} \hat{U} \bigr) \nonumber\\
&= g_{n} \mathrm{Tr} \, \hat{P}_{n} \hat{U}^{\dagger} \bigl(\sum_{m} \hat{P}_{m}\bigr) \hat{\rho}_{\mathrm{tot}}(T) \hat{U} 
- g_{n} \mathrm{Tr} \, \hat{P}_{n} \hat{U}^{\dagger} \bigl(\sum_{m} \hat{P}_{m} \hat{\rho}_{\mathrm{tot}}(T) \hat{P}_{m} \bigr) \hat{U} \nonumber\\
&=g_{n} \mathrm{Tr} \, 
\hat{P}_{n}  
\hat{U}^{\dagger} 
\left( \hat{\rho}_{\mathrm{tot}}(T) - \sum_{m} \hat{P}_{m} \hat{\rho}_{\mathrm{tot}}(T) \hat{P}_{m} \right)
\hat{U}, \label{sigmaalphafin} \\
\sum_{m}\,\sigma_{\beta} 
 &= \sum_{m} \bigl( \mathrm{Tr} \, 
\hat{P}_{n} \hat{U}^{\dagger} \hat{P}_{m} \hat{U}
\hat{P}_{n} \hat{\rho}_{\mathrm{tot}}(0)
- g_{n} \mathrm{Tr} \, \hat{P}_{n} \hat{U}^{\dagger} \hat{P}_{m} \hat{\rho}_{\mathrm{tot}}(T) \hat{U} \bigr) \nonumber\\
 &=  \mathrm{Tr} \, 
\hat{P}_{n} \hat{U}^{\dagger} \bigl( \sum_{m} \hat{P}_{m} \bigr) \hat{U}
\hat{P}_{n} \hat{\rho}_{\mathrm{tot}}(0)
- g_{n} \mathrm{Tr} \, \hat{P}_{n} \hat{U}^{\dagger} \bigl( \sum_{m} \hat{P}_{m}\bigr) \hat{\rho}_{\mathrm{tot}}(T) \hat{U} \nonumber\\
 &= \mathrm{Tr} \, \hat{P}_{n} \hat{\rho}_{\mathrm{tot}}(0) 
 - g_{n} \mathrm{Tr} \, \hat{P}_{n} \hat{\rho}_{\mathrm{tot}}(0) \nonumber\\
&= (1- g_{n}) \mathrm{Tr} \, \hat{P}_{n} \hat{\rho}_{\mathrm{tot}}(0).  \label{sigmabetafin}
\end{align}
The difference inside the parenthesis in (\ref{sigmaalphafin}) comes from the off-diagonal elements of the density matrix with respect to the degrees of freedom of the local system; 
therefore, the correction $\sigma_{\alpha}$ is the quantity related to the effect that the measurement at the final moment destroys the quantum coherence.
The correction $\sigma_{\beta}$, on the other hand, depends on the factor $g_{n}$ and independent of the state at $t=T$; 
therefore, the quantity $\sigma_{\beta}$ is related to the fact that the measured state at the initial moment differs from the eigenstate of the density matrix of the local system.
The factor $1- g_{n}$ indicates the difference from the eigenstate and $\mathrm{Tr} \, \hat{P}_{n} \hat{\rho}_{\mathrm{tot}}(0)$ is the probability of observing the state $\ket{n(0)}$.

\section{Example of an open quantum system:
One-dimensional spin chain}\label{SpinChain}

In order to evaluate the value of the correction from the microscopic reversibility quantitatively, 
we will numerically treat an open quantum system with a finite size reservoir.

\subsection{Hamiltonian and protocol}
Let us consider the total system which consists of $N$ pieces of $1/2$-spins.
We regard the first $N_{\mathrm{s}}$ spins as the local system and the remaining $N_{\mathrm{r}}(=N-N_{\mathrm{s}})$ spins as the reservoir.
The Hamiltonian reads
\begin{align}
&\hat{H}_{\mathrm{tot}}(t) = \hat{H}_{\mathrm{s}}(t) + \hat{H}_{\mathrm{r}}(t) + \hat{H}_{\mathrm{c}} \, \theta(t),\nonumber\\
\hat{H}_{\mathrm{s}}(t) &= \sum_{i=1}^{N_{\mathrm{s}}-1} J \bigl( \hat{S}^{z}_{i} \hat{S}^{z}_{i+1}
+\theta(t) \hat{S}^{x}_{i} \hat{S}^{x}_{i+1} \bigr)\\
\hat{H}_{\mathrm{r}}(t) &= \sum_{i=N_{\mathrm{s}}+1}^{N_{\mathrm{s}}+N_{\mathrm{r}}-1} J \bigl( \hat{S}^{z}_{i} \hat{S}^{z}_{i+1} + \theta(t) \hat{S}^{x}_{i} \hat{S}^{x}_{i+1} \bigr), \nonumber\\
\hat{H}_{\mathrm{c}} &= J \left( \hat{S}^{z}_{N_{\mathrm{s}}} \hat{S}^{z}_{N_{\mathrm{s}}+1}
+ \hat{S}^{x}_{N_{\mathrm{s}}} \hat{S}^{x}_{N_{\mathrm{s}}+1} \right).
\end{align}
%with
%\begin{align}
%&\hat{\rho}_{tot} = \hat{\rho}_{\mathrm{s}} \times \hat{\rho}_{\mathrm{r}} \nonumber\\
%&\hat{\rho}_{\mathrm{s}} = \exp(-\beta_{\mathrm{s}} h \sum_{i=1}^{N_{\mathrm{s}}-1} \hat{S}^{z}_{i} \hat{S}^{z}_{i+1} ) / Z_{\mathrm{s}}, \hspace{1cm}
%Z_{\mathrm{s}} = \sum_{\mathrm{config.}} \exp(-\beta_{\mathrm{s}} h \sum_{i=1}^{N_{\mathrm{s}}-1} \hat{S}^{z}_{i} \hat{S}^{z}_{i+1} ) \\
%&\hat{\rho}_{\mathrm{r}} = \exp(-\beta_{\mathrm{r}} h \sum_{i=N_{\mathrm{s}}+1}^{N_{\mathrm{s}}+N_{\mathrm{r}}-1}  \hat{S}^{z}_{i} \hat{S}^{z}_{i+1} ) / Z_{\mathrm{s}}, \hspace{1cm}
%Z_{\mathrm{r}} = \sum_{\mathrm{config.}} \exp(-\beta_{\mathrm{r}} h \sum_{i=N_{\mathrm{s}}+1}^{N_{\mathrm{s}}+N_{\mathrm{r}}-1}  \hat{S}^{z}_{i} \hat{S}^{z}_{i+1} ) \\
%\end{align}
For $t<0$, the total system consists of the two isolated Ising chains.  
We set them in the thermal equilibrium states at different inverse temperatures $\beta_{\mathrm{s}}$ and $\beta_{\mathrm{r}}$.
%Specifically, we consider the case where we connect the local system to the high temperature reservoir.
At $t=0$, we measure the energy of the local system;
the measurement basis of the initial state is the eigenstates of the density matrix of the local system, 
and thus $g_{n}=1$, $\sigma_{0}=\sigma_{\beta}=0$, and $\sigma_{T}=\sigma_{\alpha}$.
Then the local system is connected to the reservoir, so that for $t>0$, 
the total system becomes an $XY$-model. 
%We make the coupling between $N_{\mathrm{s}}$th spin and $(N_{\mathrm{s}}+1)$th spin adjustable.
%We numerically compute the break of the ``microscopic reversibility" $\sigma_{\alpha}$ 
%as we smoothly weaken the coupling between the local system and the reservoir according with the following schedule (Fig.~\ref{DecouplingSchedule}):
%\begin{align}
%h_{c}(t) = \frac{h+B}{2} -\frac{h-B}{2} \tanh[\alpha (t-\tau_{d})] \nonumber\\
%J_{c}(t) = \frac{J_{x}+B}{2} -\frac{J_{x}-B}{2} \tanh[\alpha (t-\tau_{d})] .\label{Weaken}
%\end{align}
%When $B=0$, it corresponds to the case to disconnect the system from the reservoir.
%\begin{figure}[t]
%\begin{center}
%\includegraphics[width=70mm,clip]{DecouplingSchedule.eps}
%\end{center}
%\caption{The decoupling schedule of Eqn.\,(\ref{Weaken}).
%The parameters are $h=1 (J_{x}=1)$, $\alpha=1$, $\tau_{\mathrm{d}}=5$, and $B=0.2$}
%\label{DecouplingSchedule}
%\end{figure}
Finally, we again measure the energy of the local system at $t=T$.
The strength of $J$ is spatially uniform, and thus this is the case of strong coupling between the local system and the reservoir.
This process satisfies the relation 
$\bra{m(T)} \overleftarrow{\hat{\Theta}} \hat{\rho}(T) \hat{\Theta} \ket{m(T)} = \bra{m(T)} \hat{\rho}(T) \ket{m(T)} $ 
because the Hamiltonian is invariant under the flip of all spins.
In order to calculate the time evolution, 
we diagonalize the Hamiltonian of the total system numerically with LAPACK.

\subsection{Results}
\begin{figure}[t]
%\begin{minipage}{0.5\hsize}
%\begin{center}
\includegraphics[width=80mm,clip]{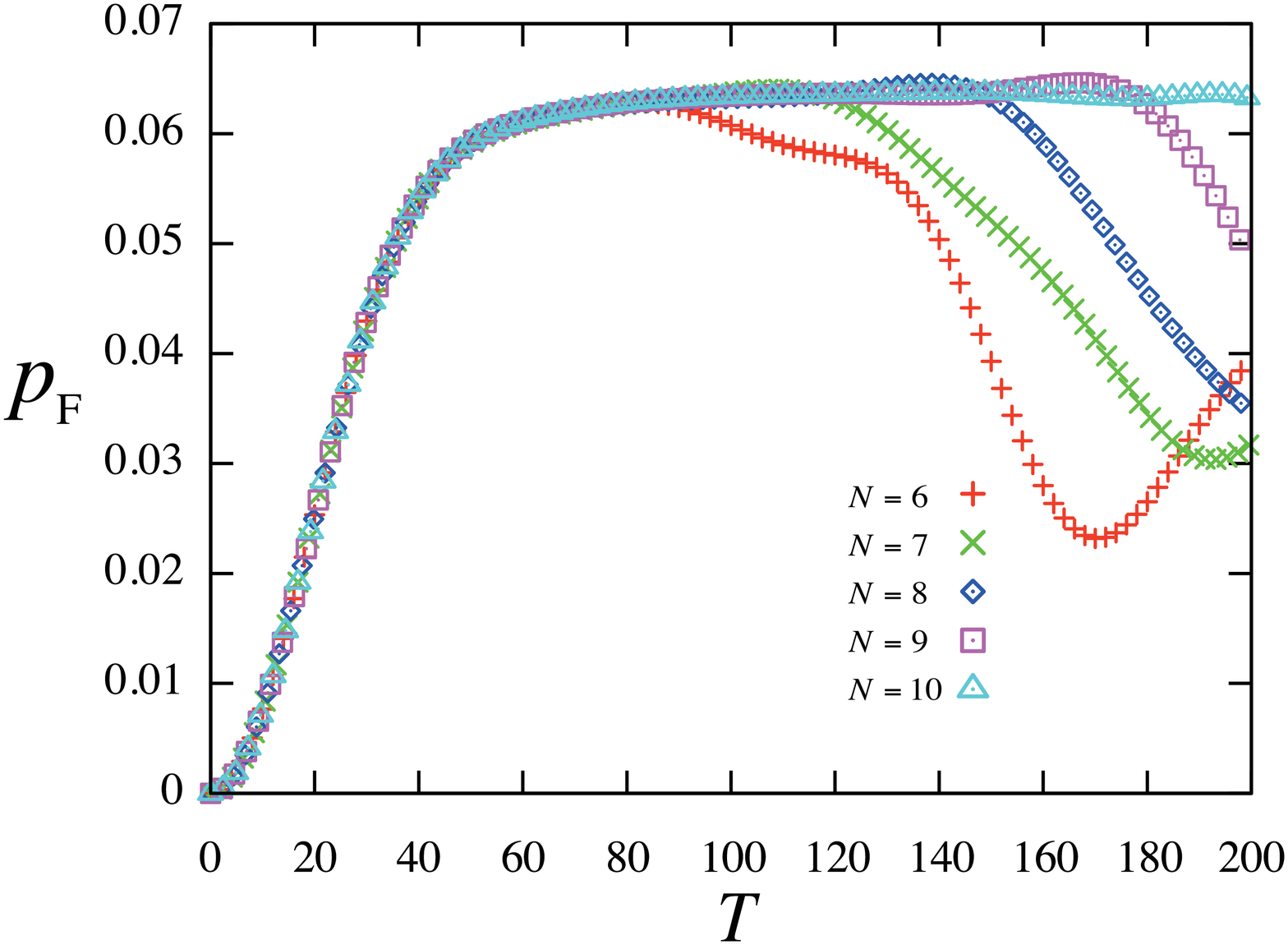}
(a)
\includegraphics[width=80mm,clip]{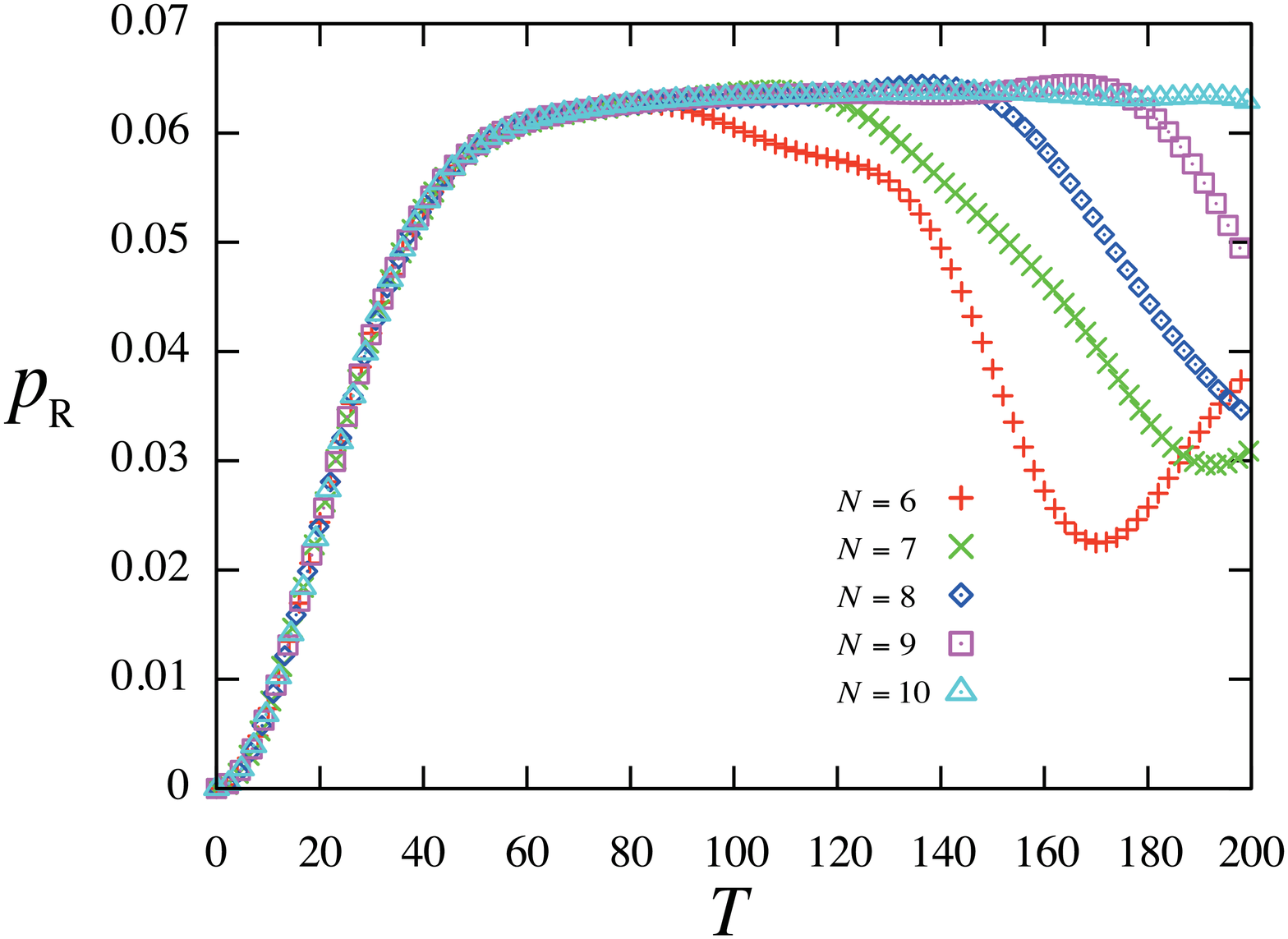}
(b)
\includegraphics[width=80mm,clip]{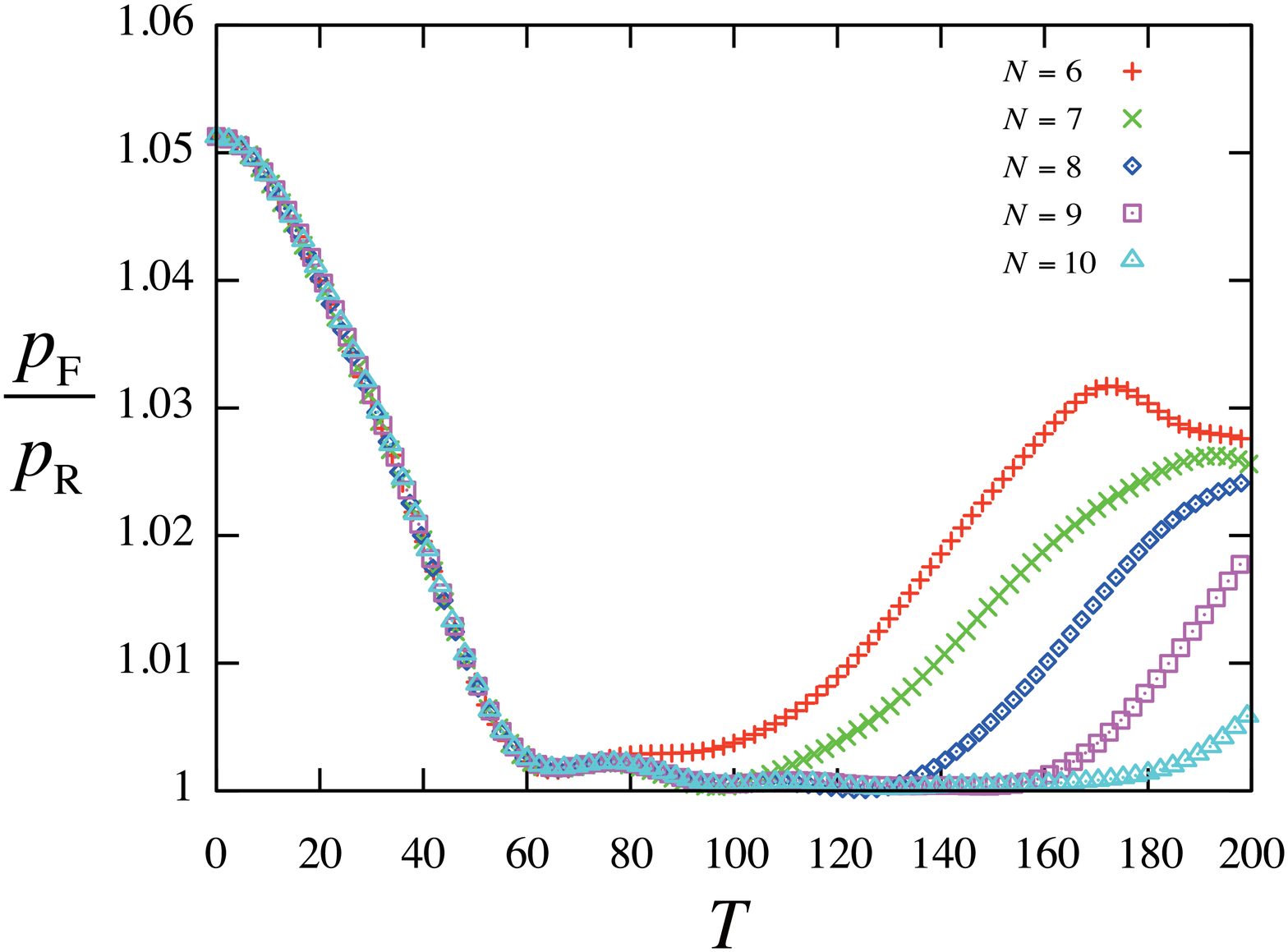}
(c)
%\end{center}
%\end{minipage}
%\begin{minipage}{0.5\hsize}
%\begin{center}
\includegraphics[width=80mm,clip]{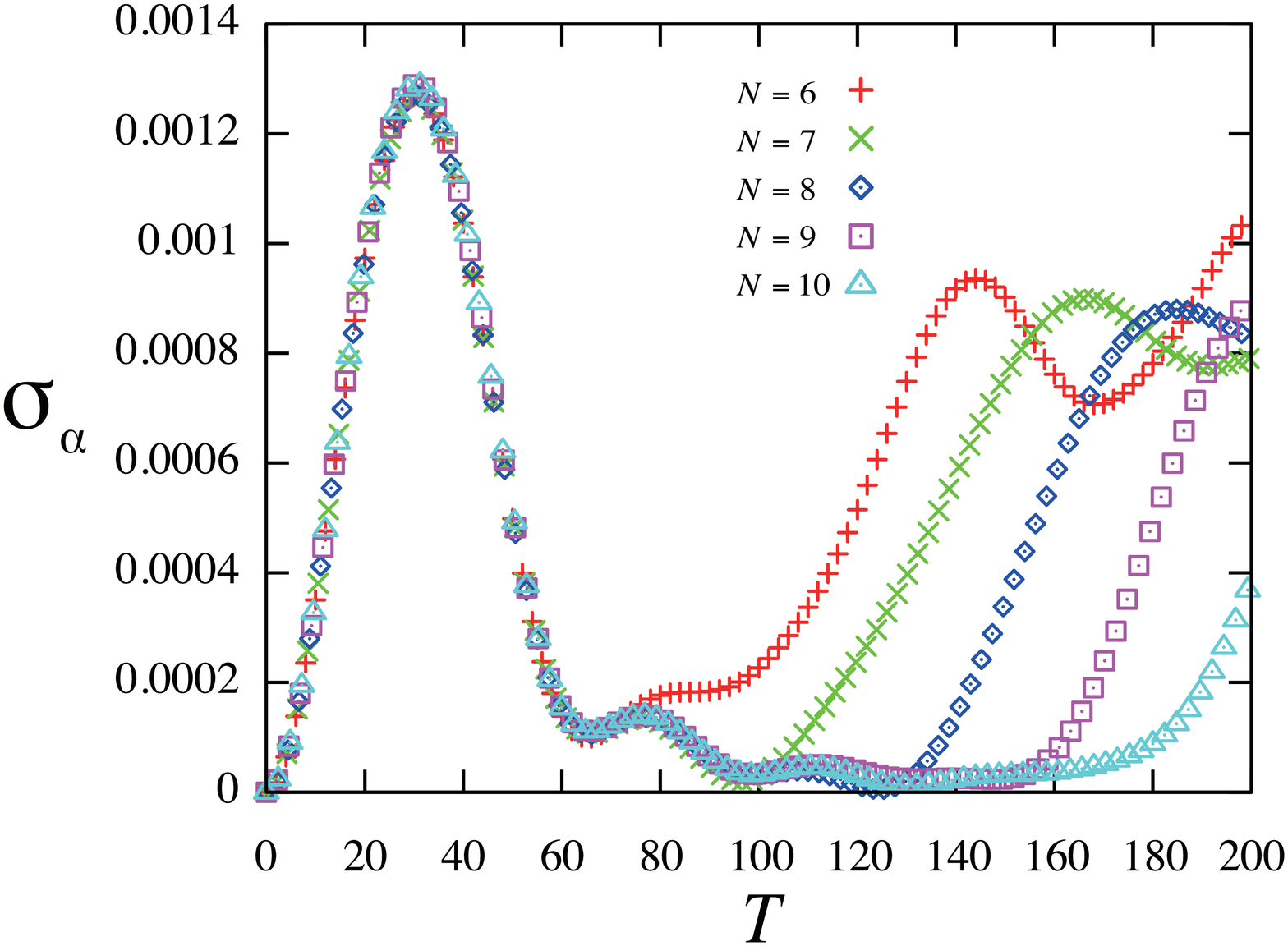}
(d)
%\end{center}
%\end{minipage}
\caption{
(Color online) 
The $T$ dependence of quantities for the spin chain with $N=6$, $7$, $8$, $9$ and $10$.  
The number of the system spins is $N_{\mathrm{s}}=2$.
We set $J=0.1$, $\beta_{\mathrm{s}}=1$, $\beta_{\mathrm{r}}=0.1$, %(\textcolor{red}{無次元量にする?}), 
and consider the transition from $\ket{n(0)}=\ket{\uparrow \, \downarrow}$ to 
$\ket{m(T)}=(\ket{\uparrow \, \uparrow} + \ket{\downarrow \, \downarrow} )/\sqrt{2}$ as the forward process.
The horizontal axis indicates the time $T$ at which we measure the local system.
The plots show the $T$ dependence of (a) the forward transition probability $p_{\mathrm{F}}(\ket{n(0)} \rightarrow \ket{m(T)} )$, 
(b) the reversed transition probability $p_{\mathrm{R}}(\hat{\Theta}\ket{ m(T)} \rightarrow \hat{\Theta}\ket{ n(0)})$, 
(c) the ratio of the forward and the reversed transition probabilities $p_{\mathrm{F}} / p_{\mathrm{R}} = 1 + \xi_{\alpha}$, 
and (d) the correction term $\sigma_{\alpha} = \xi_{\alpha} \, p_{\mathrm{R}}$.
}
\label{J01_2}
\end{figure}

We set $N_{\mathrm{s}}=2$ and varied the number of the reservoir spins $N_{\mathrm{r}}$.
Figure \ref{J01_2}a shows the $T$ dependence of the forward transition probability  $p_{\mathrm{F}}(\ket{n(0)}\rightarrow \ket{m(T)})$ 
that we obtain the measurement results $\ket{n(0)}=\ket{\uparrow \, \downarrow}$ and $\ket{m(T)}=(\ket{\uparrow \, \uparrow} + \ket{\downarrow \, \downarrow} )/\sqrt{2}$, which are the energy eigenstates of the local system, 
while Fig.~\ref{J01_2}b shows the dependence of the corresponding reversed transition probability $p_{\mathrm{R}}(\hat{\Theta}\ket{ m(T)} \rightarrow \hat{\Theta}\ket{ n(0)})$.
Figures \ref{J01_2}c and \ref{J01_2}d show the $T$ dependences of the correction terms $\xi_{\alpha}$ and $\sigma_{\alpha}$.  
We set the inverse temperatures $\beta_{\mathrm{s}}=1$ and $\beta_{\mathrm{r}}=0.1$ with the coupling strength $J=0.1$.
As long as the number of the reservoir spins $N_{\mathrm{r}}$ is finite, the finite-size effect appears as the measurement time $T$ becomes large. 
We can, however, regard the local system as an open system up to $T\sim 150$ in the simulation for $N=10$.
It shows that the forward transition probability converges to a certain value which is presumably of a new equilibrium state for  $t>0$.  
The correction term $\xi_{\alpha}$ converges to a very small (but finite) value, i.e., $\xi_{\alpha} \ll 1$.
It is rather surprising because, even though the local system approaches to an equilibrium state, the local system can be entangled with the reservoir strongly; 
the density matrix of the total system can be totally different from the product state and 
the measurement process could cause a large value of the correction term according to the discussion in Sec.~\ref{MRofOQS}.

\begin{figure}[h]
\begin{center}
\includegraphics[width=80mm,clip]{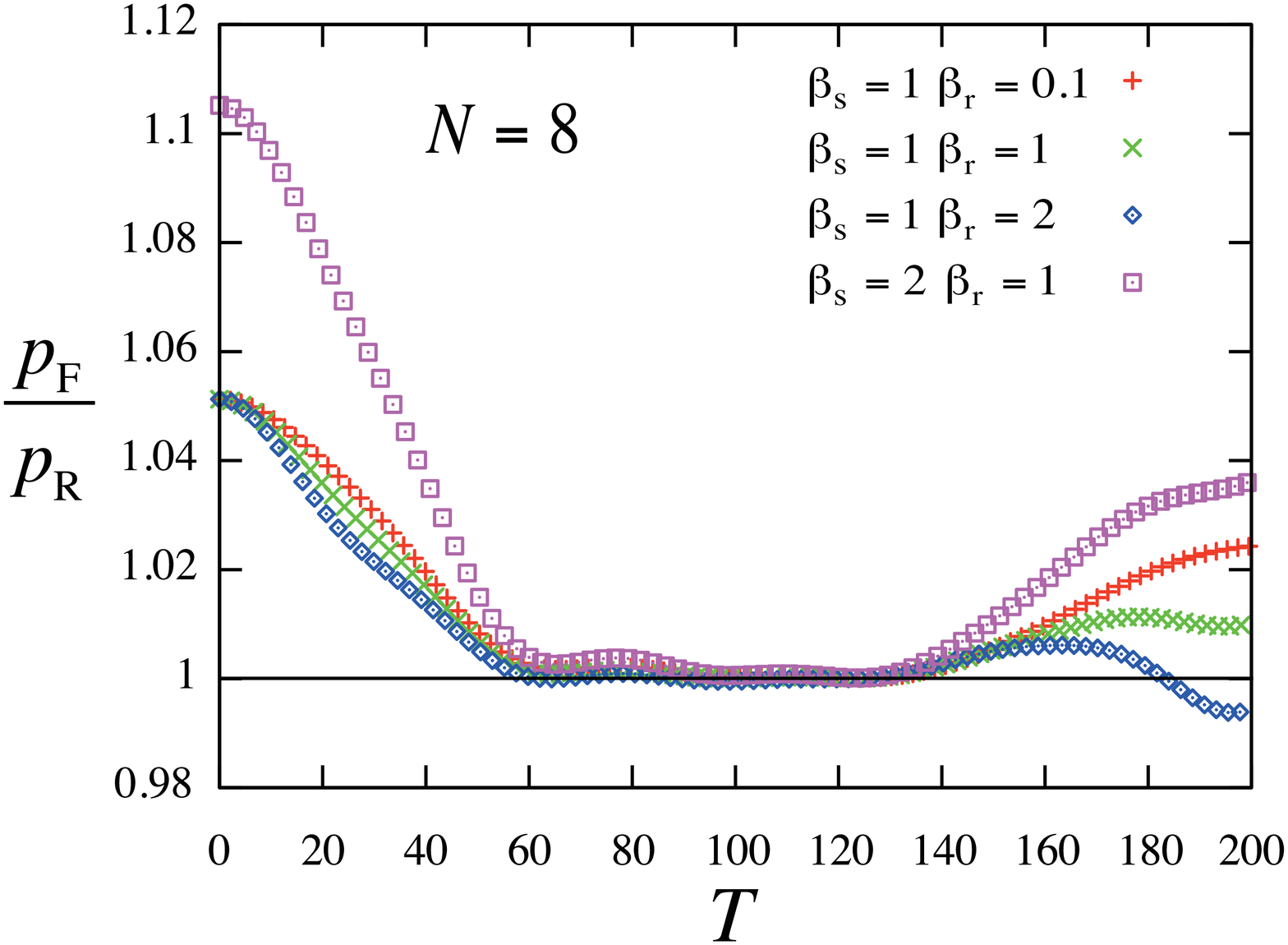}
(a)
\includegraphics[width=80mm,clip]{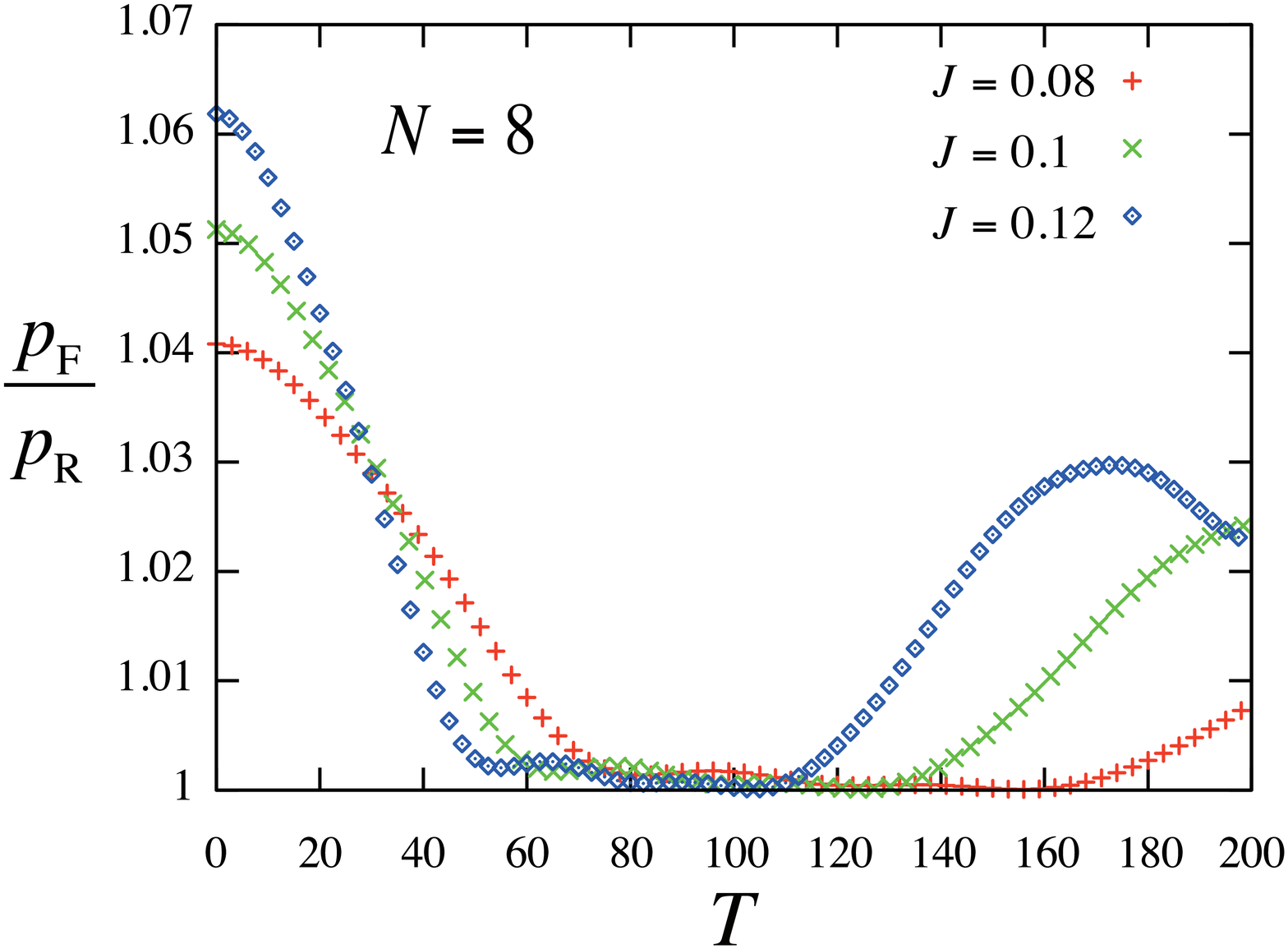}
(b)
\end{center}
\caption{
(Color online) 
The $T$ dependence of the ratio of the forward and the reversed transition probabilities 
in the case of $N=8$ when 
(a) $\beta_{\mathrm{s}}$ and $\beta_{\mathrm{r}}$ are varied with $J=0.1$ and 
(b) $J$ is varied with $\beta_{\mathrm{s}}=1$ and $\beta_{\mathrm{r}}=0.1$. 
}
\label{1_7Dependence}
\end{figure}

\begin{figure}[h]
\begin{center}
\includegraphics[width=80mm,clip]{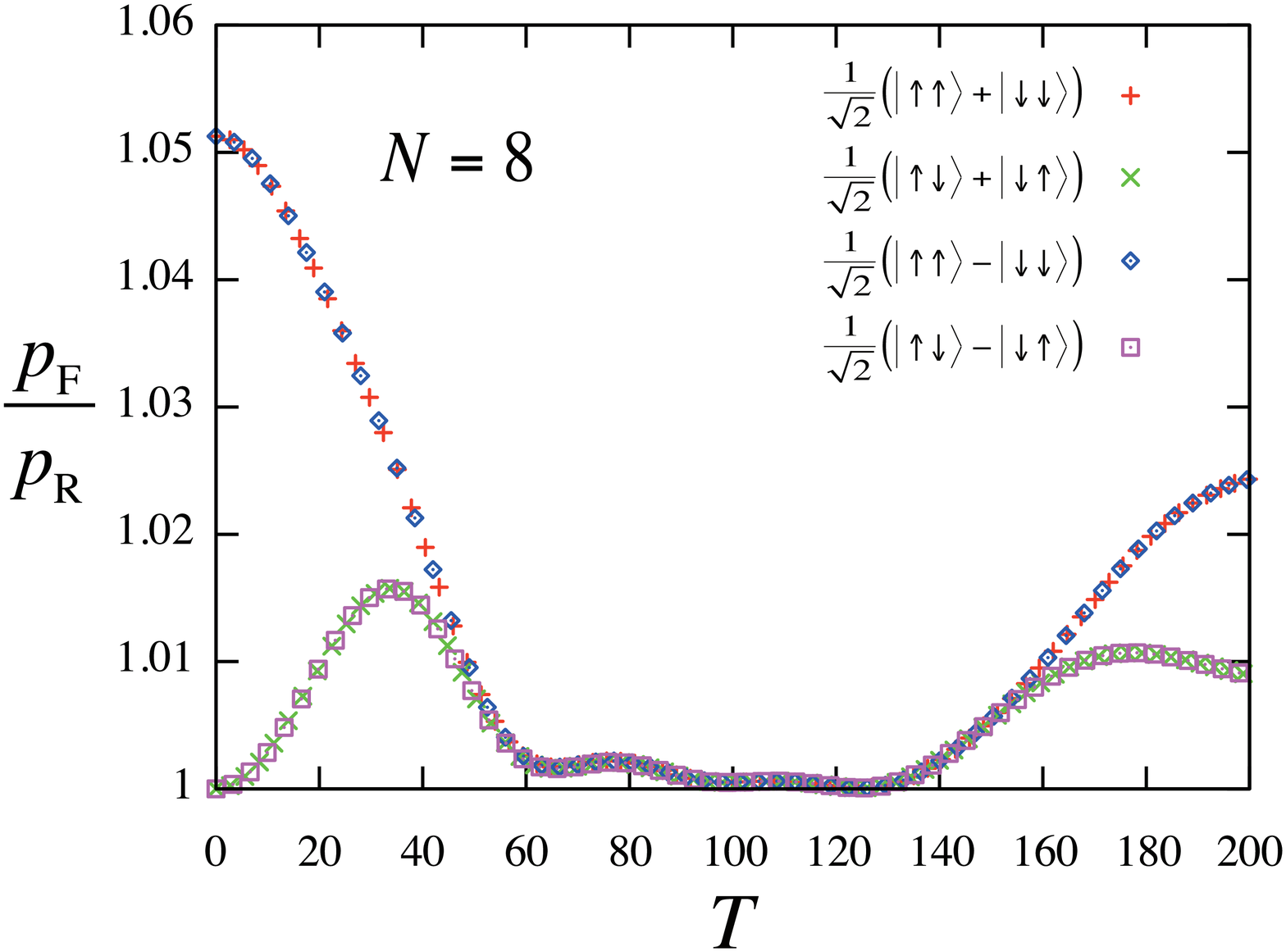}
(a)
\includegraphics[width=80mm,clip]{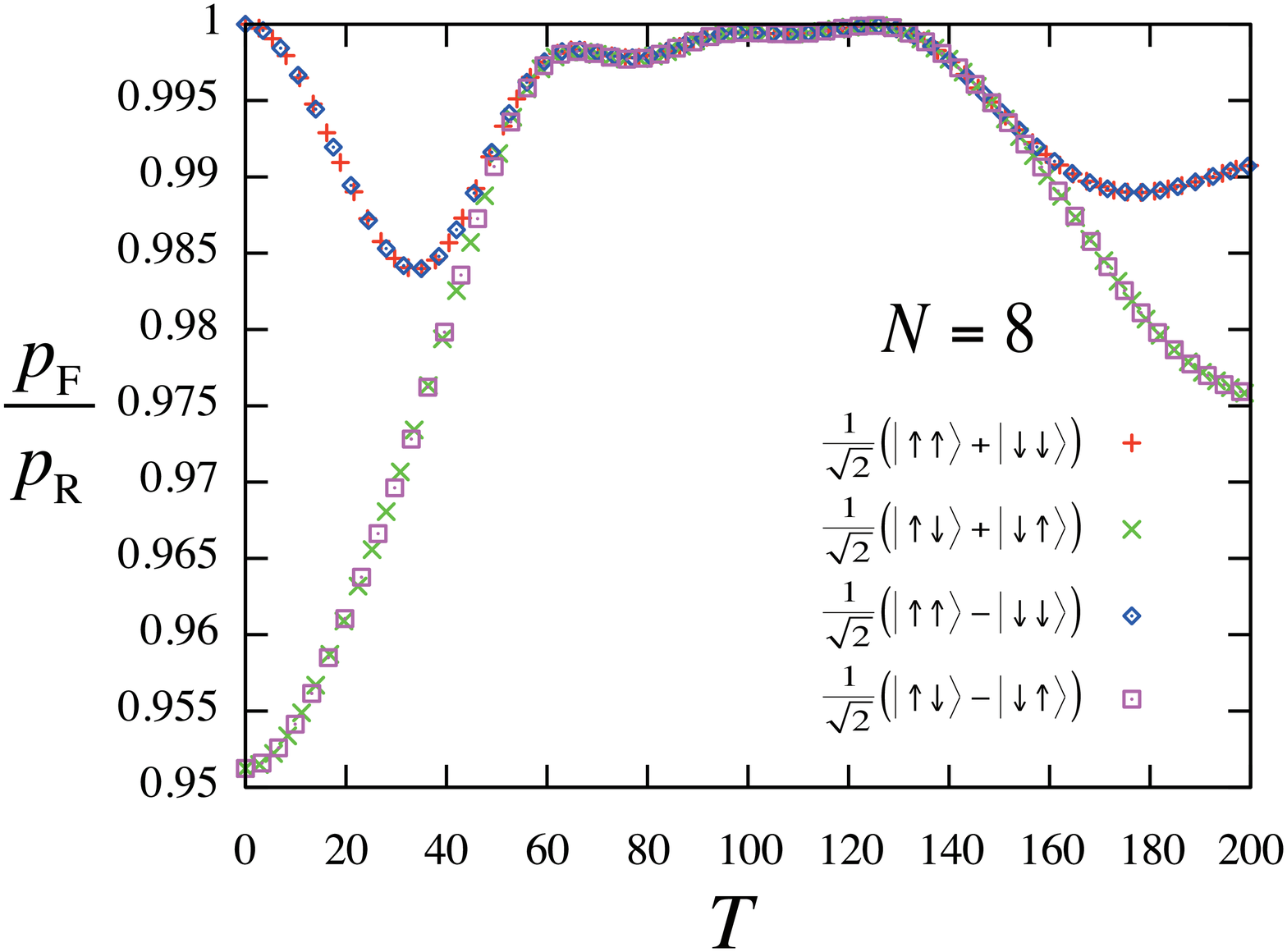}
(b)
\end{center}
\caption{
(Color online) 
The $T$ dependence of the ratio of the forward and the reversed transition probabilities 
in the case of $N=8$. 
(a) shows the result of all the transitions from $\ket{n(0)}=\ket{\uparrow \downarrow}$ and 
(b) shows the result of all the transitions from $\ket{n(0)}=\ket{\uparrow \uparrow}$.
As in Fig.~\ref{J01_2}, we set $J=0.1$, $\beta_{\mathrm{s}}=1$, and $\beta_{\mathrm{r}}=0.1$.
}
\label{1_7StateDependence}
\end{figure}

The correction seems to remain small generally in this model.
We show in Fig. \ref{1_7Dependence}a the same quantity as in Fig.~\ref{J01_2}c 
in the case of $N=8$, but by varying the value of the inverse temperatures of the local system $\beta_{\mathrm{s}}$ and the reservoir $\beta_{\mathrm{r}}$.  
For $T\sim0$, the ratio is sensitive to the value of the temperature of the local system $\beta_{\mathrm{s}}$.
The ratio 
$p_{\mathrm{F}} / p_{\mathrm{R}}$ for each parameter, however, seems to converge to a same small value as the local system goes to the new equilibrium state.
Figure \ref{1_7Dependence}b shows the same quantity as in Fig.~\ref{J01_2}c in the case of $N=8$, 
but with the coupling strength $J$ varied.
The time evolution of the system is fast for the system with a large value of $J$, 
and hence the period for which the system indicates the behavior of the open system is short. 
In the region where we can regard the evolution of the ratio 
$p_{\mathrm{F}} / p_{\mathrm{R}}$ of each parameter
as the behavior of the open quantum system 
($T\lesssim 90$ for $J=0.08$, $T\lesssim 120$ for $J=0.1$, and $T\lesssim 150$ for $J=0.12$), 
the ratios also seem to converge to a common small value as the local system goes to the new equilibrium state.
Finally, Fig.~\ref{1_7StateDependence}a and Fig.~\ref{1_7StateDependence}b show the cases of all the possible combinations of the states $\ket{n(0)}$ and $\ket{m(T)}$.
Although the sign is negative in the case of $\ket{n(0)}=\ket{\uparrow \uparrow}$, 
the absolute value of the ratio 
$p_{\mathrm{F}} / p_{\mathrm{R}}$ of 
each case seems to converge to a common small value. 
It is difficult to determine from the numerical calculations whether all these small values of the ratio 
$p_{\mathrm{F}} / p_{\mathrm{R}}$ at the stationary states coincide with each other, 
but they are of order $10^{-4}$.

Further theoretical study is required to estimate the order of the correction term compared to the transition probabilities.
Nevertheless, the present simulation suggests that we can expect that the microscopic reversibility in open quantum systems almost holds even when the local system cannot be regarded as macroscopic.

\section{Conclusion}
We derived the correction terms of the microscopic reversibility of isolated quantum systems (\ref{IsoMR1}) and (\ref{IsoMR2-1}) as well as of open quantum systems (\ref{openMR}) by formal but exact treatment.
Throughout the paper, we assumed the relation 
$\bra{m(T)} \overleftarrow{\hat{\Theta}} \hat{\rho}(T) \hat{\Theta} \ket{m(T)} = \bra{m(T)} \hat{\rho}(T) \ket{m(T)} $ 
and the product initial state for open quantum systems.  

We summarize the results of the present paper in TABLE I. 
For the microscopic reversibility in isolated quantum systems, 
we exemplified the case of a free particle system and 
found that the correction term can be very large.
For the microscopic reversibility in open quantum systems, 
we first considered two situations which seem to be physically important: 
the case where the correction terms vanish (Sec.~\ref{NoCorrections}) and the case where we disconnect the local system from the reservoir in the middle of the time evolution (Sec.~\ref{Disconnect}).
In Sec.~\ref{PhysicalMeaning}, we discussed the details of the corrections; we analyzed the bound of the factor $g_{n}$ and showed the meaning of the other correction terms explicitly by considering the quantities 
$\sum_{m} \sigma_{\alpha} = \sum_{m} \xi_{\alpha}\, p_{\mathrm{R}}$ and 
$\sum_{m} \sigma_{\beta} = \sum_{m} \xi_{\beta}\, p_{\mathrm{R}}$ 
in (\ref{sigmaalphafin}) and (\ref{sigmabetafin}).
Although we do not have an appropriate method of estimating the order of the correction terms theoretically,  
our numerical simulations of the one-dimensional spin chain suggested that, in the case of a thermal relaxation process, the correction term becomes very small compared to the transition probabilities; 
the microscopic reversibility almost holds even when the local system cannot be regarded as macroscopic.

We expect that further analyses of the microscopic reversibility reveal more interesting properties of open quantum systems.

\begin{table}
\begin{tabular}[t]{|c|c|c|c|c|}
\hline
system & initial measurement basis & final state & final measurement basis & microscopic reversibility \\
\hline
isolated & eigenstate of $\hat{\rho}(0)$ & arbitrary & eigenstate of $\hat{\rho}(T)$ & exact \\
\hline
isolated & eigenstate of $\hat{\rho}(0)$ & arbitrary & arbitrary & correction $\xi_{\alpha}$ (possibly huge) \\
\hline
isolated & arbitrary & arbitrary & arbitrary & corrections $g_{n}, \xi_{\alpha}, \xi_{\beta}$ (possibly huge) \\
\hline
open & eigenstate of $\hat{\rho}_{\mathrm{s}}(0)$ & product state & eigenstate of $\hat{\rho}_{\mathrm{s}}(T)$ & exact \\
\hline
open & arbitrary & local system disconnected& arbitrary & constant corrections\\
\hline
open & eigenstate of $\hat{\rho}_{\mathrm{s}}(0)$ & arbitrary & arbitrary & correction $\xi_{\alpha}$ (seems small)\\
\hline
open & arbitrary & arbitrary & arbitrary & corrections $g_{n}, \xi_{\alpha}, \xi_{\beta}$ \\
\hline
\end{tabular}
\caption{The microscopic reversibility for each situation.  The initial state is always assumed to be the product state for open systems.}
\end{table}

\section*{Acknowledgements}
The present author wishes to thank T. Monnai and A. Sugita for fruitful discussions and N. Hatano for useful comments.

\appendix

%\section{POVM measurement}
%If we measure the state with the POVM measurement instead of the projective measurement, 
%....

%\bibliographystyle{apsrev}
%\bibliographystyle{plain}
%\bibliography{MRinOQSBib}

\renewcommand{\refname}{\vspace{-1cm}}

\end{document}